***Attilio Sacripanti*** †§

†University of Rome II "Tor Vergata", Italy

§European Judo Union Knowledge Commission

# *Judo: the roads to Ippon*

*Biomechanics of throws tactics in competition with suggestions aimed at enhancing effectiveness of coaches tools.*

*Invited Presentation at Applicable Research in Judo Congress*

*Zagreb 02/13-14/2015*

## Abstract

In this invited lecture the biomechanical analysis of tactical tools utilized by top athletes in high level competition is presented.

In the first part, the lecture starts with a very short summary of the researches carried out on biomechanics, thermal exchange and metabolic energy expenditure related to judo throwing techniques.

These data elicit the division in more and less energetically expensive and in most and least efficient techniques in term of mechanical power applied.

Both these divisions are strictly connected to the biomechanical classification of throws in Lever and Couple.

All Judo throwing techniques: classic, innovative and chaotic are particular ways to gain some vantage by forces, in space or time, with some specific movements based on: the two physical principles underlying the throwing techniques (Lever and Couple), the athletes' body shape and structure, the referee rules, the ways to transfer energy, and so on.

The whole biomechanics of throws is analyzed concisely, pointing out the classification and singling out the most important mechanical aspects

After this introduction of the most important current researches and tendency of technical-tactical actions, the analysis of tactical attacks and the biomechanics of tools utilized, is exploited.

Researchers in the world usually consider three tactical attack modalities of the adversary: direct attack, combination and action reaction attack.

Accordingly, in this lecture, some results obtained by different researchers in each of these modalities are compared.

In the second part it is developed the specific analysis of the technical-tactical support tools.

These tools allow to enhance the effectiveness of the athletes' special throwing techniques (Tokui Waza), increasing the probability to obtain Ippon.

Besides these support tools, we consider also what coaches and athletes realize as so important: the so called ***"Technical Psychology"***: the use of some specific technical actions or new throws to increase psychological pressure and astonishment in the adversary, in order to obtain Ippon.

Biomechanical analysis allows both the classification of throws that can be combined in effective way, and the definition of how combinations should be constructed by the three groups: Chica ma waza, Ma waza, and To ma waza.

Groups that contain the most utilized throwing techniques for combinations.

The three previous classification groups arise from the variation of inter-distance between athletes.

The last part of the lecture emphasizes, in short way, the physics and biomechanics connected to judo competition that is defined as: ***“an interacting complex nonlinear system, with chaotic and fractals aspect”***.

This system must be analyzed studying the motion of “couple of athletes system” and evaluating their interaction (throws).

This conceptual simplicity highlights a very complex physical-mathematical approach.

The motion of athletes system should be analyzed by statistical Physics, whereas interaction (throws) should be analyzed by classical Newtonian Physics.

During the motion of Couple of Athletes System, human bodies show complex responses connected both to the human physiology and strong push-pull interactions.

Brownian tools are today the actual most sophisticated way of modeling motion in competition, starting from fractals till to multi-fractals aspects.

However not all is easy in the analysis of throwing techniques, from the physical point of view, except for the basic principles Couple and Lever.

If we study the specific mechanics of several throwing action in real competitions we can face some interesting physical aspects not totally well known, like: almost-plastic collision between extended bodies and for some Lever techniques we have to solve dynamical problem of bodies with variable mass that differs from the well known Classical Mechanics with constant mass.

In effect, in such a situation, the general methodological approach of mechanics must be properly modified; considering a deformable body with a finite extension, where special emphasis is given to the variable rotational inertia during throw, with interesting consequences about balance, linear and rotational momentum and motion of center-of-mass.

In this lecture all tools, both well known at coaching like level and new proposals as well, are classified and organized in scientific way, stressing the biomechanical principles that rule their application in high level competitions and assessing these basic tools to enhance effectiveness.

The clarification and understanding of the inner mechanics of these tools is a powerful help for coaching to teach competition tactics in useful way.

***Attilio Sacripanti***

# *Judo: the roads to Ippon*

*Biomechanics of throws tactics in competition with suggestions aimed at enhancing effectiveness of coaches tools*

*Invited Presentation at Applicable Research in Judo Congress*

*Zagreb 02/13-14/2015*

*Attilio Sacripanti*

# *Judo: the roads to Ippon*

## *Biomechanics of throws tactics in competition with suggestions aimed at enhancing effectiveness of coaches tools*

## 1. Introduction

Ippon is the final goal of judo athletes and coaches.
The part of judo, that best approximates the ideal of aesthetic beauty of Dr. Jigoro Kano.
Timing, speed, strength, power, coordination and control are the basic ingredients of this aesthetic expression, which fascinates every viewer has the world experts or inexperienced.
However defensive capabilities are so increased that it is very difficult to obtain Ippon, with a simple attack, without the addition of some specific tactical tools which increase the effectiveness.
According to Japanese method, competitive tactics is expressed in Renzoku and Renraku waza, but researchers have classified three main types of tactical actions: direct attack, combinations and action reaction attack.
In this article in detail the mechanics of Judo throwing techniques will be analyzed and thanks to biomechanics, also tools widely used on world tatami will be identified.
Will be analyzed individually the three types of tactic classified by the researchers, the better to be able to compare the data of the other analysis of tactics in high level competitions, in the end will be identified specific tools utilized.
A short overview of the physics behind motion and interaction in judo will close this article to better show the complexity hidden by the apparent simplicity of the solutions.

## 2. Biomechanics of throwing techniques

The biomechanics of the Judo throws was widely analyzed in many previous papers till from 1987 [1, 2, 3]. Both from the Biomechanical and Thermal-Energetic and Metabolic point of view. [4,5].

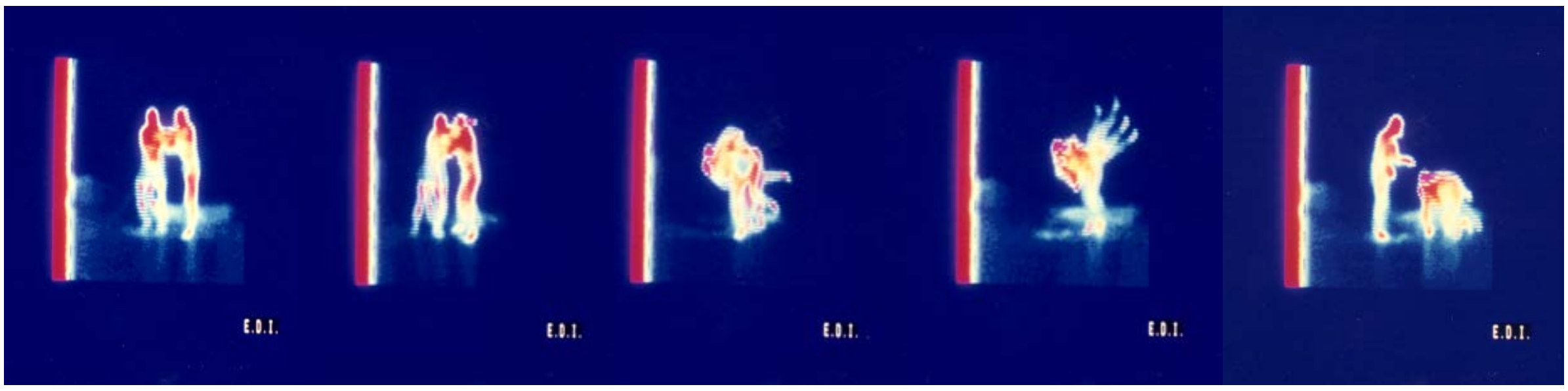

***Fig.1-5. Thermo-grams of Koshi Guruma***

Verifying the well known, old and first in the world Japanese study and results (1958) [6] that Lever techniques are more expensive in term of metabolic energy consumption, than Couple techniques.

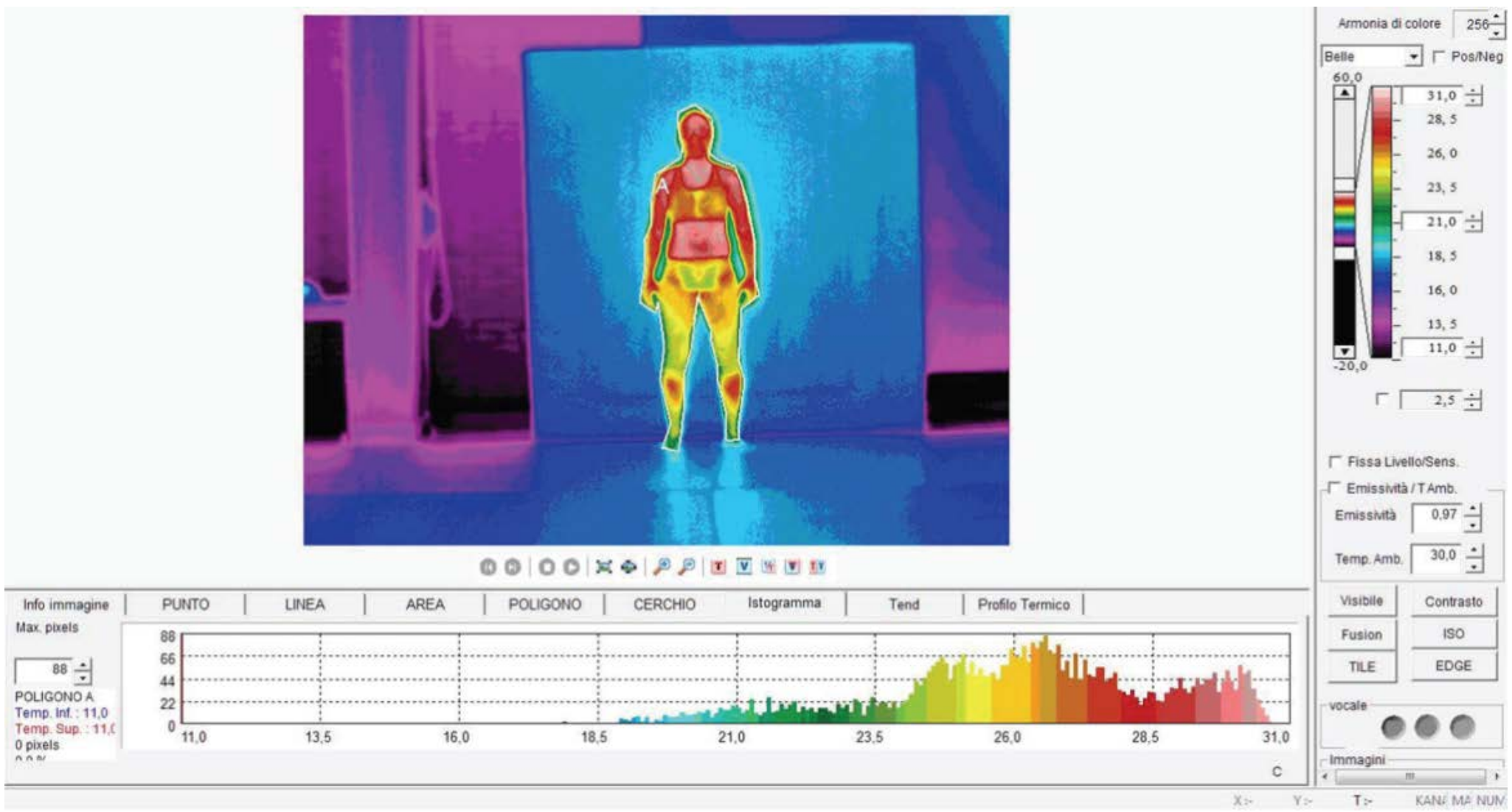

*Fig 6 Final Thermo-gram after Nage Komi* **[5]**

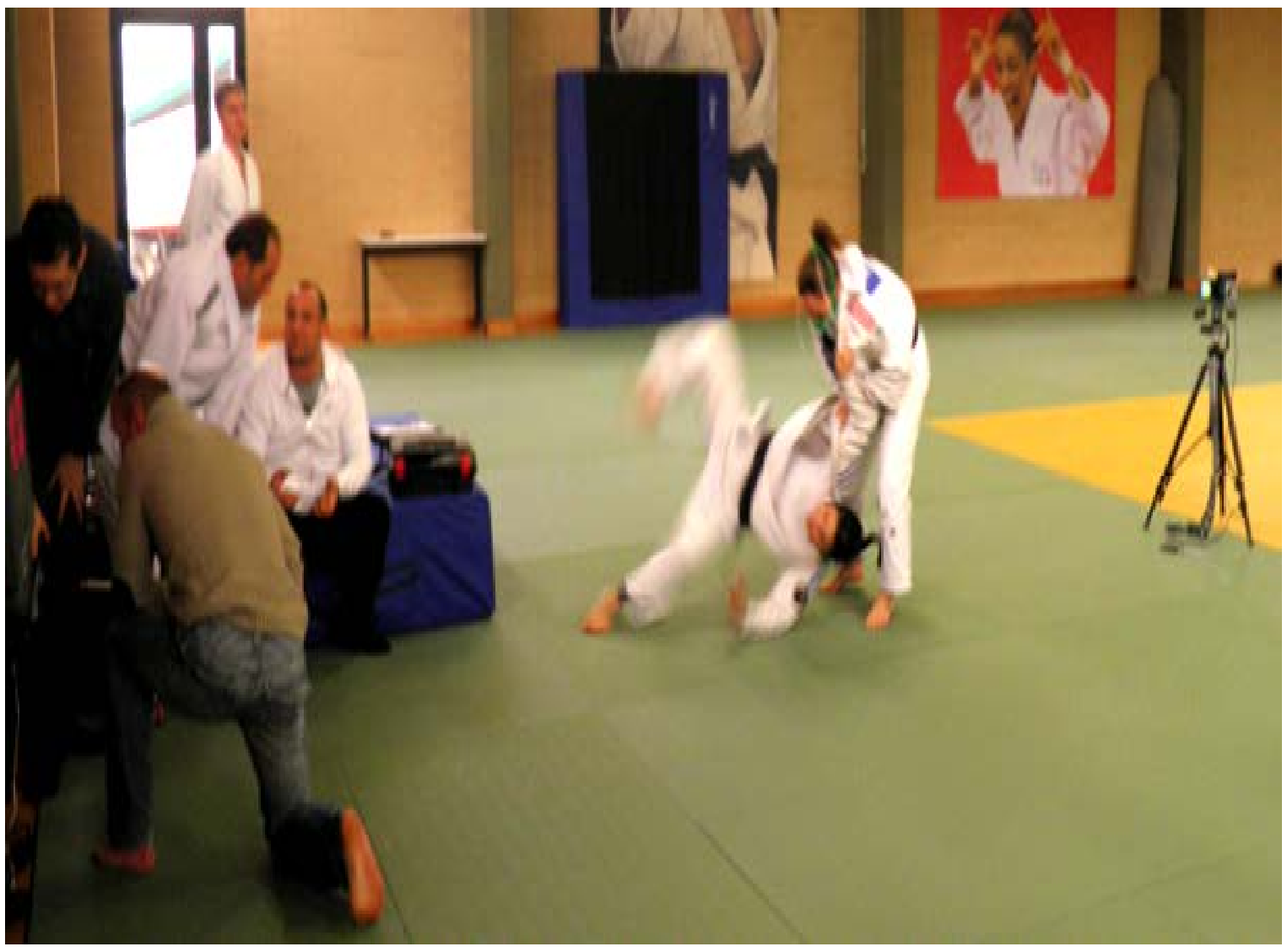

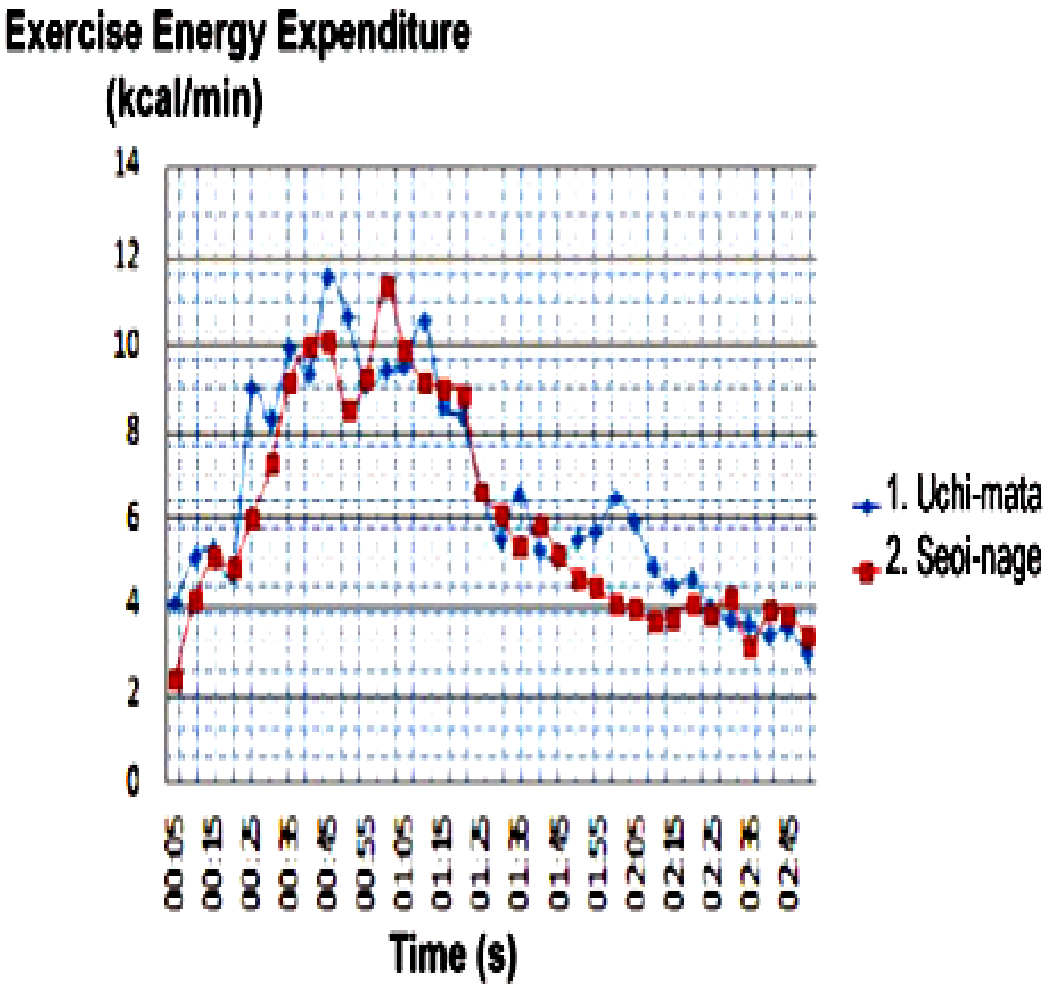

*a)* *b)*

***Fig 7.***

***a) Set-up for capturing both infrared thermal images and metabolic energy consumption during nage-komi of lever- vs. couple-throws*** [5]

b) ***Graphic representation of Energy expenditure of two different throws ( lever-vs-couple) in nage-komi (graph to the right).*** [5]

These finding are supported by biomechanical evaluation of the whole movement of these two group, in fact Lever techniques are on the average built by more complex movements than Couple techniques, and also intrinsic techniques mechanics underlines that, because Couple developing a rotation around CM are less expensive than Lever who are based on the translation of CM in space. [7] Some results are shown in the following figure and tables.

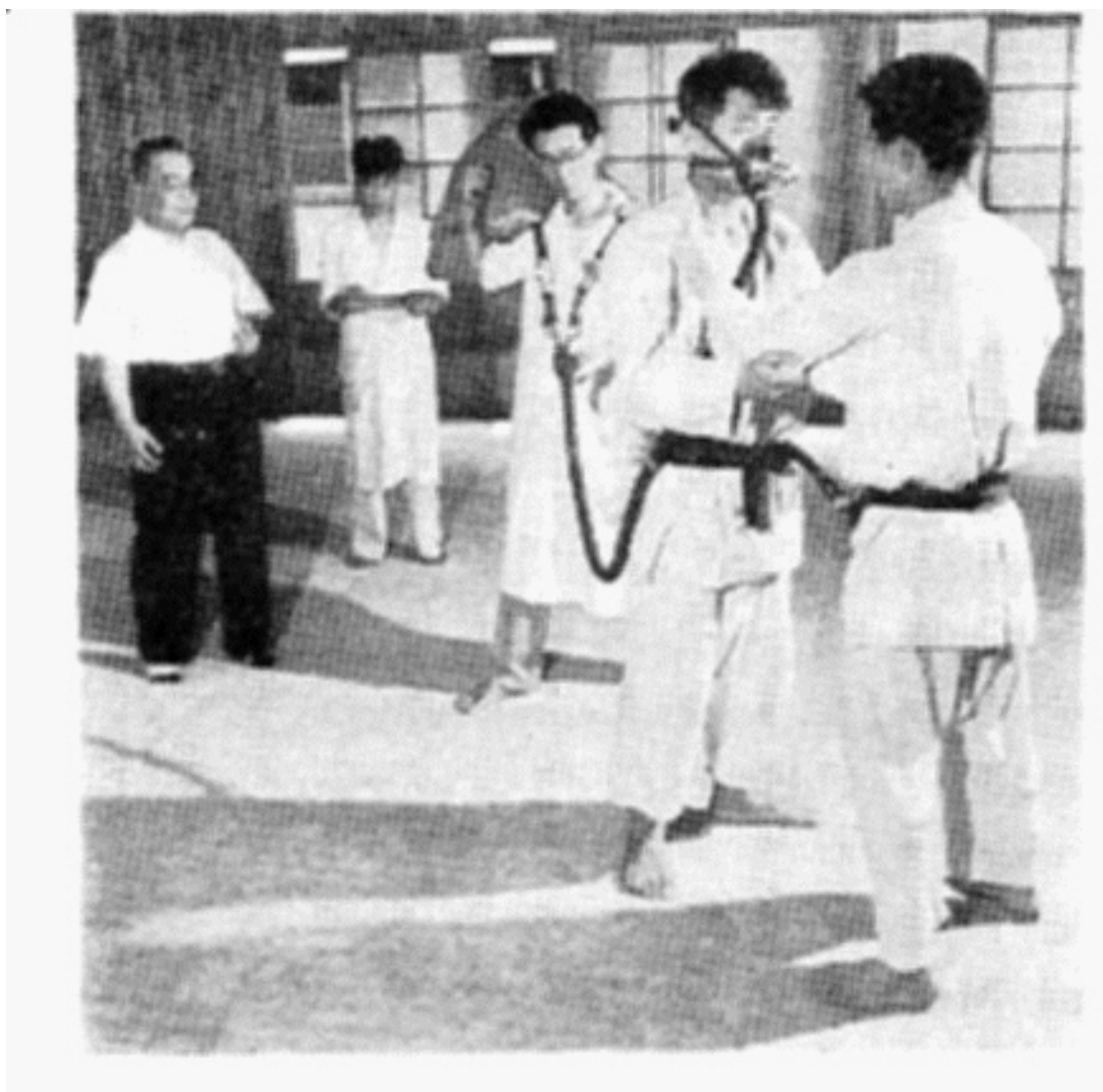

| Techniques | Average of offensive | Average of defensive |
|---|---|---|
| Defensive attitude | 18.2 | 13.5 |
| Uchimata | 2.9 | 2.6 |
| Osotogari | 3.5 | 2.7 |
| Ouchigari | 4.0 | 2.8 |
| Sutemiwaza | 3.6 | 3.5 |
| Seoinage | 4.0 | 3.5 |
| Ashiharai | 2.9 | 2.4 |
| Hanegoshi | 4.3 | 3.4 |
| Attitude directly prior to Newaza | 18.5 | 17.7 |
| Yokoshihokatame | 18.4 | 30.2 |
| Kuzurekamishiho-katame | 21.0 | 29.9 |

***Fig 8 tab 1 first Japanese experiment on metabolic energy expenditure of judo throws [6] (1958)***

| **Tools** | **Judo -technique** | **K Joule** |
|---|---|---|
| Co | Uchi Mata | 4,5 |
| u | Ashi Arai | 3,6 |
| ple | O Soto Gari | 4,8 |
| Le | Ippon Seoi Nage | 6,3 |
| ve | Koshi Guruma | 6,8 |
| r | Tai Otoshi | 5,7 |

***Table 2 Energy differences between couple and Lever techniques ( unpublished data 1990)***

| Throwing: ↓ | Peaks → S (N) | Peaks → P (W) | * | Time in milliseconds to reach the peak: ↓ | |
|---|---|---|---|---|---|
| | | | | F | P |
| **Ippon-Seoi-Nage** | **702,9** | **166,5** | * | **767** | **1480** |
| **Uchi-Mata** | **543,2** | **272,8** | * | **1106** | **1468** |

***Tab 3  Efficiency of  Throws   [8]   Couple is more efficient  than  Lever***

In the previous  tab. 3  are shown the results obtained  by Peixoto and  Monteiro , on the efficiency comparison  between  Couple and Lever .

Results show that   Ippon-Seoi-Nage   reaches a higher (702.9N) and earlier (767ms) peak of strength but a lower peak of power (166.5W).

Uchi-Mata  reaches a lower (543.2N) and  later (1106ms) peak of strength but reaches a  higher peak of power(272.8W) .showing higher efficiency than the lever throws.

Summarizing the judo throwing techniques structure is based, simply, on two physical tools applied by Tori on Uke. To apply these two tools is essential the contact between bodies. Then judo throws are structured as shortening distance followed by a collision plus application of one tool. The shortening of distance between the athletes into the Couple of Athletes System is solved by the geodesic between athletes. Only three trajectories (for all throwing techniques) the shortest and fastest one’s, called  GAI (General Action Invariant) are normally used before collision. [7]    Fig. 9, 10, 11, 12.

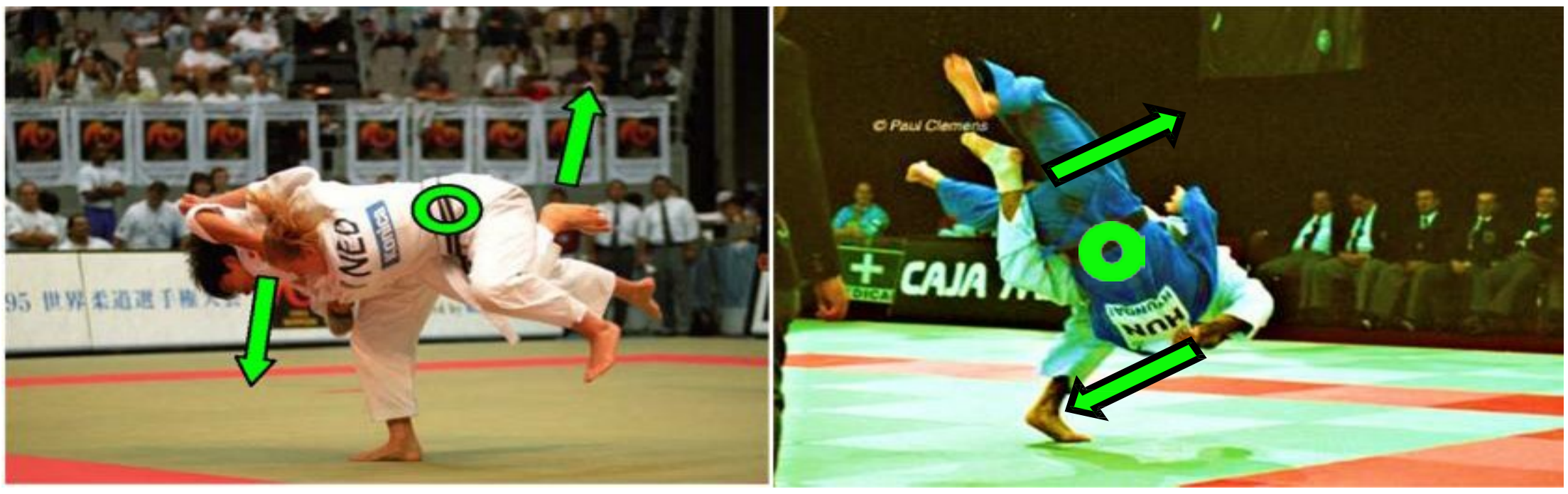

a)

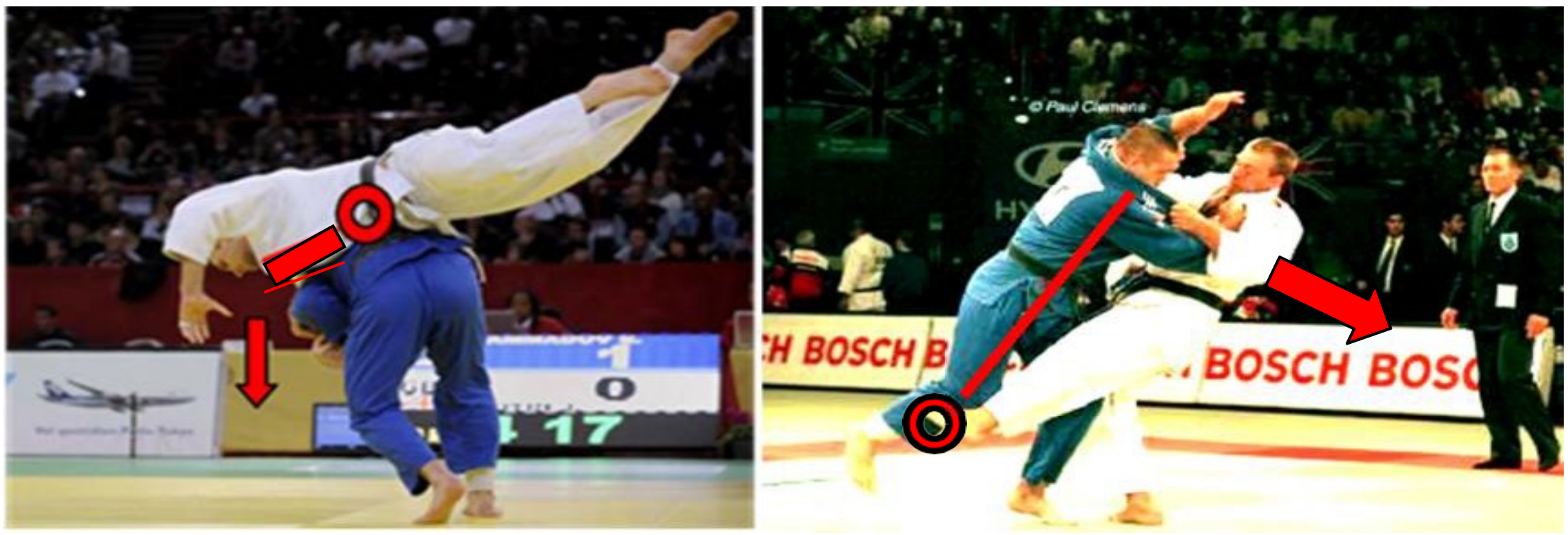

b)

***Figg.9, 10, 11, 12  Throwing in competition,***
***a)   Application of a couple of  forces        and   b) Application of a physical lever***

It is possible also to use different and longer trajectories, however GAI are the most energetically convenient. After these common phases (GAI+ Collision+ Tool [Couple or Lever]) Judo throwing techniques split in two main ways, depending by the physical tool utilized.

1. If the athlete applies as tool a Couple, only timing is important, there is no other needs, for example (theoretically) no unbalance, no stopping time and so on.
2. If the athlete applies as tool a Lever, he needs to finish off the technical action by a series of secondary musts: stopping the adversary motion for a while, unbalancing him, positioning of the fulcrum and harms action (application of forces in right direction).

For the second one, the musts are connected to the superior and inferior kinetic chains motion and right positioning connected to Kuzushi and Tsukuri phases.

Consequently, it is possible to resume the basic steps (useful for coaching and teaching) of jūdō interaction (throwing techniques) that occur during Dynamic phases of high level Competitions.

These steps really reflect a continuous fluent movement:

1. *First: breaking the Symmetry to slow down the opponent (i.e., starting the unbalancing action)*
2. *Second: timing, i.e., applying the "General Action Invariant", with simultaneously overcoming the opponent's defensive grips resistance,*
3. *Third: sharp collision of bodies (i.e., the end of unbalance action)*
4. *Fourth:*
    A. *Application of "Couple of Forces" tool without any need of further unbalance action,*
    B. *Use of the appropriate "Specific Action Invariants", needing to increase unbalance action, stopping the adversary for a while, applying the "Lever" tool in Classic, Innovative or Chaotic way.*[9]

These steps represent the simplest movements which occur to throw the opponent.

Very often though, far more complex situations can arise under real fighting conditions. These complex situations which have evolved from the simple steps explained above depend on the combination of attack and defensive skills of both athletes.

However, the actual *Collision* step is very important for applying any real throwing technique

It is important to underline that many countries have made and are still making new technical contributions to *Kōdōkan* jūdō.

The motivation towards victory and the ability to overcome opponents represent the root of this evolution (without implying either a positive or pejorative meaning, and merely using the term in a sense of "changes over time"). Some examples of these newly added techniques entries are: the ***Korean seoi-nage,*** the ***Cuban –sode-tsuri-komi-goshi,*** the ***Korean sukui-nage***, the ***Russian gyaku-uchi-mata*** by Shota Chochosvili or the ***Khabarelli***, which was a well-known "free wrestling" technique

(*cfr.* "*Advances in judo biomechanics research*"[10]), or the ***Armbar*** by Neil Adams, to name just a few. These techniques cannot be found in most standard jūdō books.

The main analyst of the Innovative aspect of throws is ***Roy Inman*** from England [11-14]. These kinds of new throwing variations very often arise from the will to apply classic techniques in highly changeable dynamic situations. In doing so, their users may be able to catch their opponent off-guard with such high dynamic throws.

From a biomechanical point of view, it is important to define and understand how such techniques may be developed. Therefore, it is possible in terms of biomechanical analysis conclude that:

***Classical Throws: All the throwing movements as showed in the Kano's Go Kyo 1922, and Kodokan's Go Kyo 1985*** [15].

***"Innovative Throws" are all throwing techniques that keep alive the formal aspect of classic Jūdō throws, and differ in terms of grips and direction of applied forces only.***

In our definition, ***Innovative Throws*** are ***henka*** (variations) applications of classical *Kōdōkan* throwing techniques, which biomechanically are either Couple of Force-type or Lever-type techniques, while it remains easy to still recognize a basic traditional Classic technique (40 *gokyō* throwing techniques) in them.
However, there are other "non-rational or non-classic" solutions applied in competition and which are different form 'Innovative' (*henka* Throws), which we define as ***New*** or "***Chaotic Forms***".
The feeling of 'different' is produced by chaotic non-identifiable throwing techniques from a view point of those accustomed to seeing classic *Kōdōkan gokyō* expressions of throws.
Throwing techniques produced in such way are normally called 'new' or Chaotic present a number of difficulties in facing them.

***Chaotic Throws principally belong to the Lever-type Group, and are characterized by the application of different grips positions (SSAI) which applying force in different (nontraditional) directions while simultaneously applying (ISAI) in non-classical positions, with also non rational take-contact trajectories (GAI) or Couple applications.***

Normally Chaotic forms of throws are special tricks prepared in advance to astonish the adversary in specific important competition ( Olympic Final or topic moment during fight) normally are utilized as technical Psychological tool to win.
Their effectiveness is very high because the astonishment stops most of defensive capabilities of the adversary.

It is possible to summarize all the biomechanical phases of judo throwing techniques both in static or dynamic situation, into a simple global diagram, see Diag.1

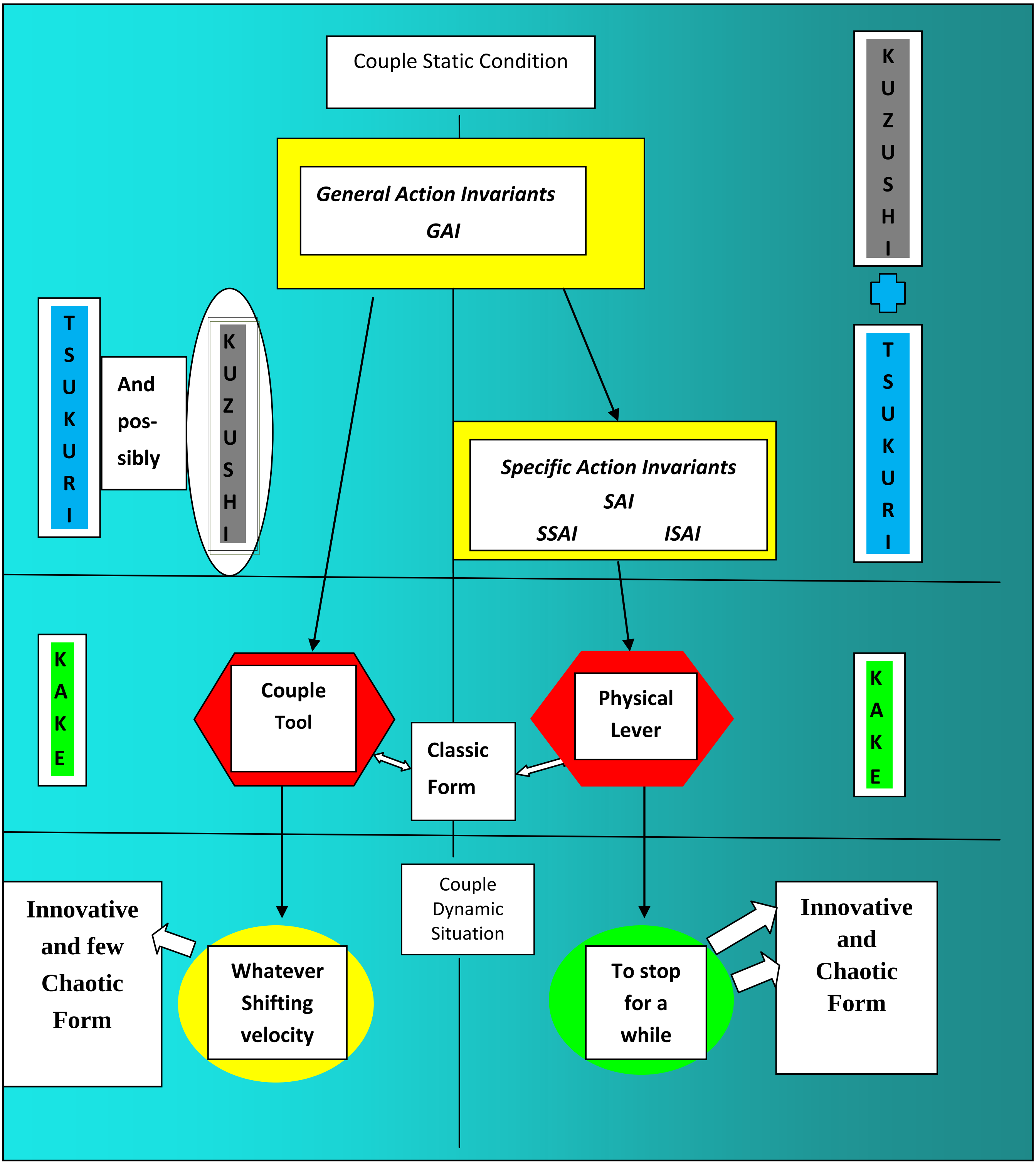

*Diagram 1: Summary of the Kuzushi Tsukuri Action Invariants connected to Kake phase for: Classic, Innovative and Chaotic Form of throwing techniques* [9]

Then in function to both physical tools all judo throwing techniques (also the not allowed today by referee regulation) can be grouped and classified as in the following tables.4,5

<table>
<tr>
<td rowspan="4">Techniques<br>Of<br>Physical lever<br><br>Lever applied with</td>
<td>Minimum Arm<br>( fulcrum under<br>Uke’s waist)</td>
<td>O Guruma Ura Nage<br>Kata Guruma Ganseki Otoshi<br>Tama Guruma Uchi Makikomi<br>Obi Otoshi Soto Makikomi<br>Tawara gaeshi Momo Guruma<br>Makikomi Kata sode Ashi Tsuri.<br>Sukui Nage Daki Sutemi<br>Ushiro Goshi Utsuri Goshi</td>
<td rowspan="4">All<br>Innovative<br>Variation<br>and<br>New<br>(Chaotic)<br>Forms</td>
</tr>
<tr>
<td>Medium Arm<br>( fulcrum under<br>Uke’s Knees)</td>
<td>Hiza Guruma<br>Ashi Guruma<br>Hiza Soto Muso<br>Soto Kibisu Gaeshi</td>
</tr>
<tr>
<td>Maximum Arm<br>( fulcrum under<br>Uke’s malleola)</td>
<td>Uki Otoshi Yoko Guruma<br>Yoko Otoshi Yoko Wakare<br>Sumi Otoshi Seoi Otoshi<br>Suwari Otoshi Hiza Seoi<br>Waki Otoshi Obi Seoi<br>Tani Otoshi Suso Seoi<br>Tai Otoshi Suwari Seoi<br>Dai Sharin Hiza Tai Otoshi<br>Ikkomi Gaeshi Tomoe nage<br>Sumi Geshi Ryo Ashi Tomoe<br>Yoko Kata Guruma Yoko Tomoe<br>Uki Waza Sasae Tsurikomi Ashi<br>Uke Nage</td>
</tr>
<tr>
<td>Variable Arm<br>(variable fulcrum<br>from the waist<br>To the knees)</td>
<td>TsuriKomiGoshi O Goshi<br>SasaeTsurikomi Goshi Koshi Guruma<br>Ko Tsurikomi Goshi Kubi Nage<br>O Tsurikomi Goshi Seoi Nage<br>Sode tsurikomi Goshi Eri Seoi Nage<br>Uki Goshi Morote Seoi nage</td>
</tr>
</table>

***Tab.4.Techniques based on a lever*** [9].

<table>
<tr><td rowspan="5">Techniques<br>Of<br>Couple of forces<br><br>Couple applied by</td><td>Arms</td><td>Kuchiki daoshi Uchi Mata Sukashi<br>Kibisu Gaeshi<br>Kakato Gaeshi<br>Te Guruma</td><td rowspan="5">All<br>Innovative<br>Variation<br>And<br>Very few<br>Chaotic form</td></tr>
<tr><td>Arm/s and leg</td><td>De Ashi Barai O Uchi Gari<br>Okuri Ashi Barai Ko Uchi Gake<br>Ko Uchi Barai Ko Soto Gake<br>O Uchi Barai Harai Tsurikomi Ashi<br>Tsubame Gaeshi Yoko Gake<br>Ko Uchi gari O Soto Gake<br>Ko Soto Gari O Uchi Gake</td></tr>
<tr><td>Trunk and legs</td><td>O Soto Gari O Tsubushi<br>O Soto gruruma O Soto Otoshi<br>Uchi Mata Ko Uchi Sutemi<br>Okurikomi Uchi Mata Harai Makikomi<br>Harai Goshi Ushiro Uchi Mata<br>Ushiro Hiza Ura Nage<br>Hane Goshi Gyaku Uchi Mata<br>Hane Makikomi Daki Ko Soto Gake<br>Yama Arashi ( Khabarelli)</td></tr>
<tr><td>Trunk and arms</td><td>Morote Gari</td></tr>
<tr><td>Legs</td><td>Kani Basami</td></tr>
</table>

***Tab.5. Techniques based on a couple*** [9]

As it easy to see, techniques flowing from Couple tool are more or less half of the techniques flowing from Lever tool, this because biomechanically speaking the Couple throws are simplest than the Lever throws .
Tori's body motions are both toneless and easier; some throws are applied by same body movement having the same biomechanical essence, even if Japanese called them in different way, treating them as different things.
Japanese way to name these movements (as we see in Kazuzo Kudo (工藤和三) [16]) is useless ; because, hiding the true physical basis, it is grounded on movement memory only.
In fact if physical tools are understood by athletes, application, whatever grips or direction can arise, would be quite easy and natural.
Therefore if it is well understood how to apply Couple tool, then whatever is the hand position in grips the movement is only one, application of Couple into the three plane of symmetry; in application of these techniques timing is complex and very important.
Instead for Lever tool, whatever is forces direction by grips, one can apply stopping point in many positions, despite that, throwing mechanics is always the same ( To turn the adversary' s body around the stopping point).
In these techniques timing is less important but complex coordination of arms, legs and body is most important.

## 3. Judo Competition Studies

Competition is the main area of throwing techniques application, and is the focus of the coaching and scientific analysis and researches.
For coaches competition is defined by many models one of these is the well known ***"Hajime-Matte Model"*** this model stands for the basic technical steps between a start and stop in Competition, and it is arranged at maximum in six phases: **Mobility, Kumi Kata, Preparation, Nage Waza, Transition, Ne Waza.** [16]
For Scientists also there are many Competition models, for example in term of energy consumption, or as action time development, or even more sophisticated computer models, but for biomechanics in spite of a difficult definition: ***"Judo competition is an Interacting complex nonlinear system, with chaotic and fractals aspect"***, the phases analyzed in contrast with the coaching model are only two: **Locomotion** (of the Couple of Athletes System) and **Interaction** (Throws or Ne Waza techniques).
However this formal simplicity hides a very complex mathematical approach for the "simple" motion of the athletes' couple system.
From the other side, interaction (throws) that seems more complex, may be analyzed by means of "the simple" Classical Newtonian Physics.
The main argument of this paper will be about the biomechanical tools utilized for throwing techniques enhancement.
About studies on Competition, a lot of advanced researches have been performed on the connection between grips and throws in different condition or with different adversaries [ 18, 19, 20, 21, 22. 23] or grips and tactics [24].
Also difference in throwing techniques utilized by junior and senior, or men and women [ 25, 26, 27]
Very few researches were produced with the quality of practical coaching indications or utilization.
Among these one of the most interesting is the group of researches performed in France, by Calmet and coworkers "*Optimisation de la performance en judo: analyse des combats de haut-niveau*" [28].
Sertic, on the other hand, showed how personal and broad is the judo expression in high level competition, underlining the needs to train by individual approach or by homogeneous groups.
Emphasizing that this approach could enable bigger efficiency and greater chance to achieve good international results.[29]
One paper analyzed also among other variables the biomechanical effectiveness of throws during London Olympics [30] see tab.6. Heinish presented at EJU Poster exhibition 2014 [31] a deep and long study (2004-2013) on high level competition analyzing the tendencies and efficacy on technical-tactical action in judo showing the increasing tendencies of ashi waza and inward rotation techniques. Drid performed one analysis of Olympic games in London 2012 [32] highlight that movement structure of judo sport is considered to be highly demanding for the majority of judokas muscle groups; therefore high level of strength, power, speed and flexibility are essential for success in judo. Segedi and Sertic analyzed the effectiveness and importance of throwing techniques in judo Matches, by a questionnaire to eight Judo experts; classifying all judo techniques by Importance in Two groups less and most Important connected to weight categories [33] Some results are shown in the next table 7.

| *Throws Effectiveness In London Olympic 2012* | | |
|---|---|---|
| ***Throws*** | ***Effectiveness Male %*** | ***Effectiveness Female %*** |
| Seoi ( Ippon – Morote - Eri) | 14.8 (329) | 8.2 (222) |
| Uchi Mata | 9.2 (138) | 15 (143) |
| O Uchi Gari | 15 (53) | 24 (49) |
| Ko Uchi Gari | 12 (57) | 37 (35) |
| Tai Otoshi | 25 (36) | 23.8 (21) |
| Soto Makikomi | 10 (10) | 23.6 (17) |
| Tani Otoshi | 46 (13) | 50 (16) |
| Uchi Mata sukashi | 90 (10) | 100 (10) |
| ***Couple*** | ***28.7*** | ***39*** |
| ***Lever*** | ***24*** | ***26.4*** |

***Tab.6 Throws Effectiveness In London Olympic 2012*** **[**30]

| *Classification by Importance of throwing techniques for weight categories in Judo Matches* | |
|---|---|
| B2 block –***Lever group***<br>***Light -Medium weights*** | B3 block- ***Couple group***<br>***All weights*** |
| C22 -Tomoe Nage | C32- O Uchi Gari |
| C23- Sode Tsurikomi Goshi | C34 – Uchi Mata |
| C24- Morote Seoi Nage | C35- Ko Uchi Gari |
| C25- Ippon Seoi Nage | |

***Tab.7 Throws Effectiveness by Weight Categories*** **[**33]

Other field of interest connected to tactics in competition, was the study of utilization of time during contest, these studies were connected to difference in throws used by junior and senior, or to tactics of attack or pace of attack in high level competition [34, 35].
Tactics in competition was the main argument for many studies, starting from the historic work of Sterkowicz and Maslej on the throw tactics in high level competitions [36] not the first but very complete, till to the recent work of Miarka, Calmet and Boscolo del Vecchio [37] a complete and analytical review of techniques and tactics in judo, from 1996 to 2008.
In these studies analysis of tactic is grounded on a comparison of frequency distribution of events over a range of factors such as: time structure, number of applied techniques and directions, number of successful actions per minute, quality of attacks awarded with points, movements of elite judo athletes development, and grip forms.
In other papers, tactics is connected to the main athletes attack ways: ***Direct attack, Action-Reaction strategy*** and ***Combination of Throws*** these are the main ways to attack the adversary in high level competition.

Analytical results are connected to tournament, some show that the Action–reaction methods are the most useful , followed by Combination and Direct attack. [38] Some other show that Direct attack is the most utilized and effective, and so on.
In the world Championship 2010 happen in Japan the following table shows the same use of the same different attack types [39].Tab.8

| ***Judo Attack Type inWorld championship 2010 Japan*** | |
|---|---|
| AttackTactics | % |
| DirectAttack | 42.2 |
| ActionReaction | 34.8 |
| Counter Attack | 16.3 |
| Combination | 8,1 |

***Tab.8. Percentage of different types of attack WC Japan 2010*** [39].

It is interesting to see the Japanese fighting Style utilized during the same world championship in the following table clearly appears that Japanese athletes with little change prefer the dynamic style couple techniques more than lever, with an interesting supplement, more sutemi to overcome the fighting style of foreign people see Tab.9

| ***Judo Skill of Japanese Male team WC 2010 Japan*** | |
|---|---|
| Techniques | % |
| Ashi Waza | 33.3 |
| Te Waza | 13.3 |
| Koshi Waza | 6.6 |
| Sutemi Waza | 33.3 |
| Ne Waza | 12.6 |

***Tab.9 Judo Skill of Japanese Male team WC 2010 Japan*** [39].

In recent times researchers look more interested in connection among grips configuration and throws applied.
The approaches and kumi-kata type (grip form) give way to specific behaviors between the two contestants, but again we specify that it is the whole body position into the Couple of athletes system , the so called ***"Competition Invariants"*** that drive the throw effectiveness.[9]
Studies by Courel and co-workers [40] tray to identify the effects of grip laterality and throwing side combinations (i.e., attacking on the same side of the gripping, or vice versa) on attack effectiveness and combat result in elite judo athletes.
The results were that attacking on the same side of the kumi-kata increase the chance of scoring and winning the combat independently of sex and weight category.
Perform same-side attacks by kenka-yotsu (adversaries using reverse grip, ex: right versus left) it was the most effective one, especially for lightest weight judo fighters.
Both perform same-side attacks by ai-yotsu using right or left grip at the same time; and only one athlete gripping (only the athlete attacking performing the grip);are situations more open and increases the likelihood of winning the combat for both opponents.

Recent study [41] compares the throwing techniques efficiency index performed from the same and opposite grip for senior male and female during Bosnia and Herzegovina senior State championships. Male dominate in regards to the same side grip, while female dominate in regards to the opposite side grip.
The most efficient throwing technique for male considering the same side grip was as expected Ippon seoi nage, (Lever) while for the female Harai goshi (Couple).
The highest efficiency technical index both for male and female regarding the opposite side grip was for Uchi mata (Couple)
Some new data about connection between grip configurations and throwing techniques are presented in [42]. This last area of studies is clearly connected to the increase of physiological pressure, by technical tools, like grips.

## 4. "Technical Psychology" and Tactics in competition

**What is Tactics**?

The Encyclopedia Britannica defines it: ***"in warfare, the art and science of fighting battles on land, on sea, and in the air. It is concerned with the approach to combat; the disposition of troops and other personalities; the use made of various arms, ships, or aircraft; and the execution of movements for attack or defense."***
Translating for judo: ***"The art and science of fighting in judo competition; it concerns the approach to combat: the study of grips related to relative bodies' positions, the use of advantageous displacements, the way to open defense, the change in velocity, and the execution of technical movements for attack or defense."***
In Biomechanical term: ***"The science to rule and manage body's mass, angular momentum and kinetic energy, using in right way forces in space and time to throws or hold the adversary in a Judo contest."***
In the fields of previous definition, Biomechanics can help strongly not only to understand and clarify the various points but also to single out the tools useful for the throwing enhancement.
The technical tools it is the well known "Match Analysis" worldwide utilized heavily by the National Federations [43].
Athletes are not only complex mechanical machine, but also (strong or weak) psychological systems, and psychology can play a very important part in grab the victory or lose the way.
It is well known that sport psychology is an interdisciplinary science that draws on knowledge from many related fields like: biomechanics, physiology, kinesiology and psychology.
Psychological issues play a major role in optimizing health and improving performance of athletes in judo.
Psychological principles are important to understand and maintain motivation for grueling training regimens over long periods of time, dealing with pre-competition preparation and stress during competition, and redefining goals and objectives after competition. [44]
Coaches and athletes at all levels realize also the importance of ***"Technical Psychology"*** which is a very narrow branch of this science that is grounded ***on the disturbing pressure and surprise action by means of pure technical tools,*** like: positions, grips, specific paces, utilization of referee rules, new Tokui Waza. Chaotic forms of throws, and so on.
Many years ago with different refereeing rules the technical psychology was fixed to the combat at the edge of fighting area, many time increased by kenka yotsu. [18]

Today with the change in the rules, at first level (disturbing situation) of this psychological approach it is found the kenka youtsu grip position, in which athletes is put under pressure by an opposite grips.
But the key factor of this approach is the surprise produced by construction of new not known throwing techniques.
These techniques, that are the main way to use “technical psychology”, are the so called ***New*** or ***Chaotic Form of Throwing Techniques,*** normally the astonished athletes are caught out and oppose weak defense in these situations.
To increase drastically their effectiveness, Chaotic throws are upgraded by the main biomechanical tools utilized to enhance effectiveness in tactical attacks.
The first among these techniques and the first application of technical psychology in such specific way it is found in the historic and well known “***Khabareli Throw***” that the Georgian Champion Shota Khabareli as Soviet Union, gold medal at Olympic 1980, used during his athletic life.
See the next figures 13, 14.

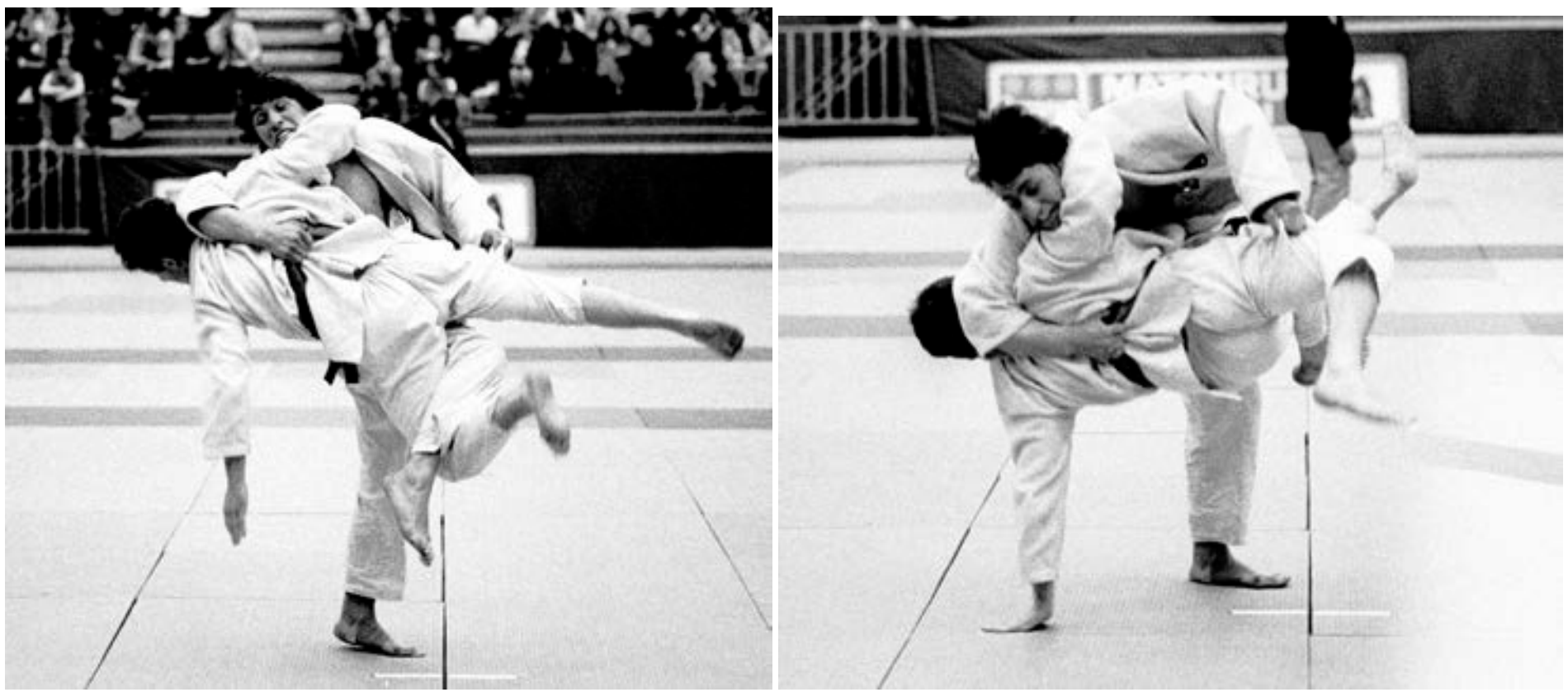

***Fig13,14- Khabarelli: master of unorthodox grip and attack –This throw it is not o uchi gari changed in a strange lifting… (Adams); but it is a well known free wrestling throw called backward Kliket made more easy by the belt and trousers grips, how it is possible to see in the illustration of Prof Petrov’s book. ( picture Finch)***

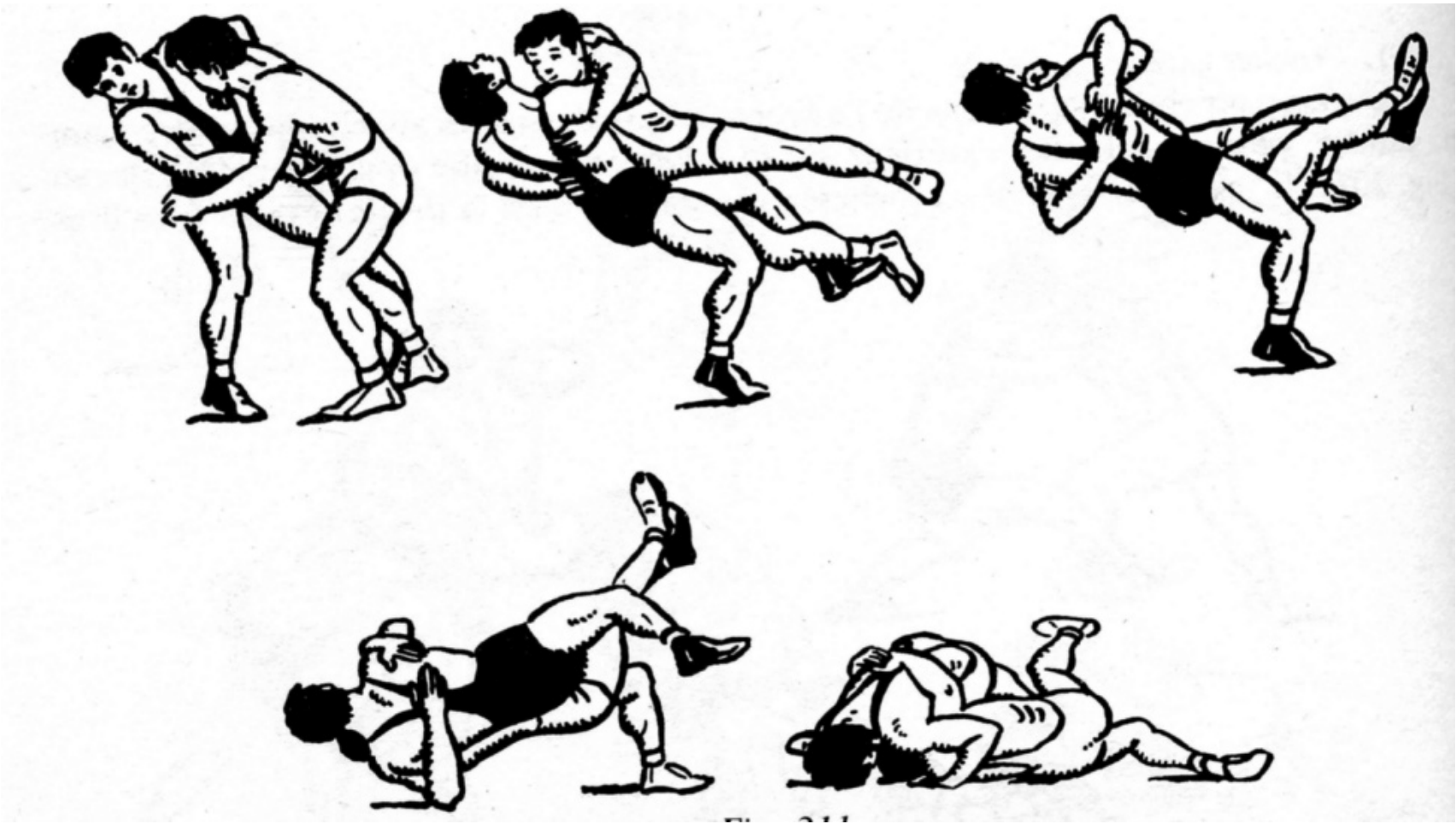

***Fig.15 -Backward Kliket***
***From Rajko Petrov Lutte Libre et Lutte Greco Romaine***

In the following Fig. 16-19 there are shown four example of Chaotic throwing techniques, two of Lever and two of Couple, applied once in a while in high level competition, as psychological tricks with high effectiveness.

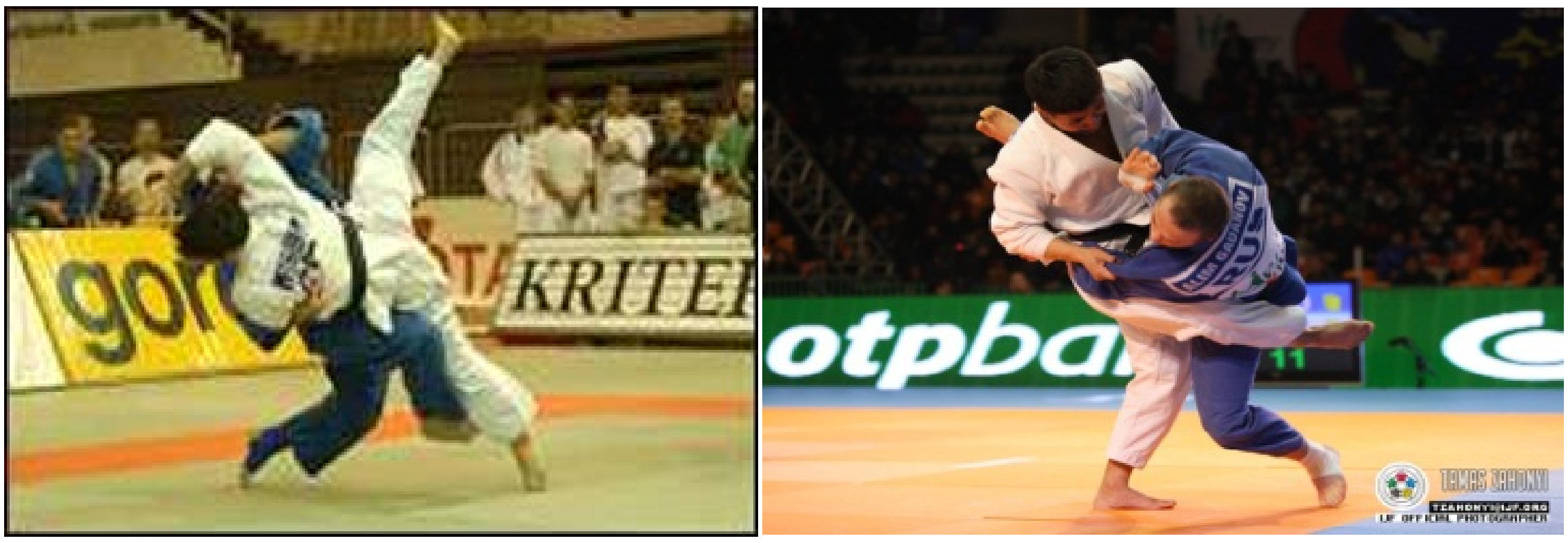

a)

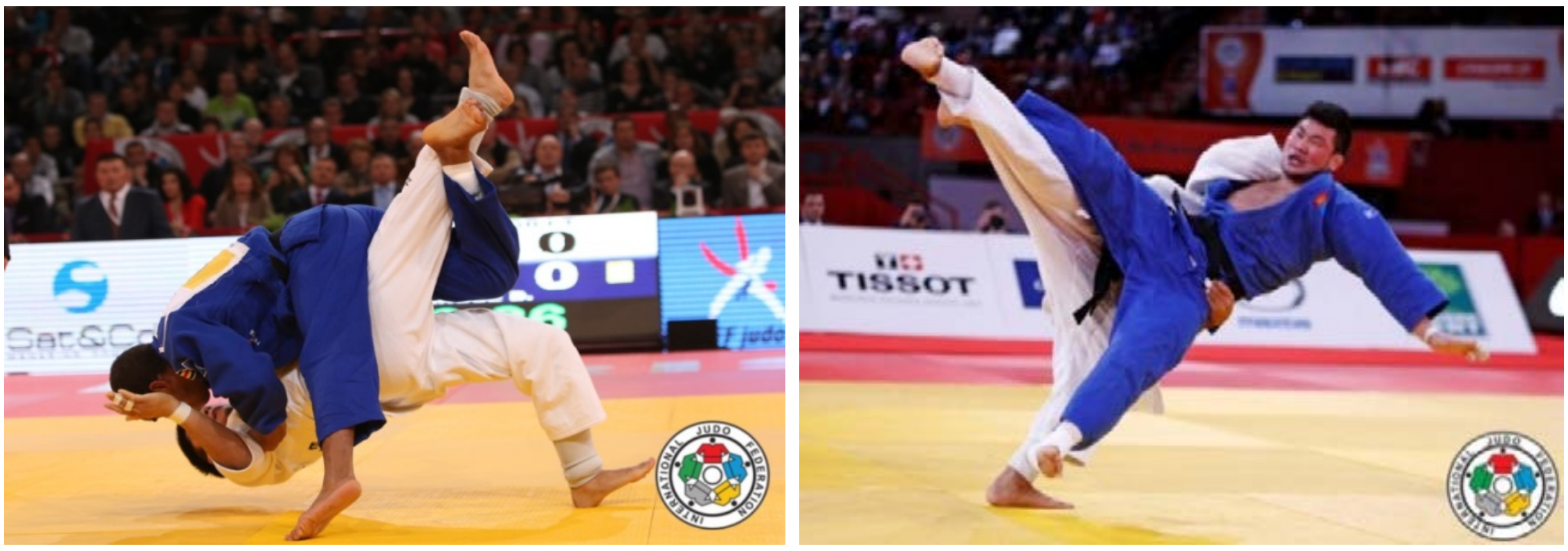

b)

***Fig 16-19 Chaotic Form all these techniques are application of the two basic biomechanical principles of throws but in very different form from Kano classified Technique, with high technical-psychological impact and effective results.***

***a) Lever application     b) Couple application***

## 5. Tools in direct attack

For many years tools to enhance throwing effectiveness were based both on attack time decreasing (this means shorter attack trajectories, ex.: two steps against three steps) and weakness in human body structure exploitation (this means diagonal attack along the unable in defense body sides).
Today a third element is combined in this enhancement, the utilization of rotational dynamic approach. Obviously many coaches already solved this problem but in this paper it is clarified in term of biomechanics.
This also because the practical mechanical solution in term of trajectories, movements and forces directions follow the rules found by Newton in his classical mechanics.
The biomechanical analysis performed try to find the basic paradigm of the problem, to help and improve the technical training focalized on the effectiveness of throws in competition.
Then it is possible to meet a rotational action pre-post added to Couple techniques; and new technical variation based on rotational unbalance action in case of lever techniques.
The results gathered are summarized in easy and simple steps, as follow.
Couple Techniques are enhanced in their effectiveness in three ways:

1. The Couple tool that, in principle, doesn't need unbalance allowing uke's body rotate around his center of mass, it is enhanced utilizing the Uke's body smaller resistance directions (normally summarized in Diagonal attacks).
2. The vertical rotational movements in the transverse plan with the axis in the sagittal plane can be added to ( post) the Couple application with Transverse Rotation , and axis in the frontal plane to overcome some defensive resistance, mainly in the trunk and leg group of Couple techniques (like Uchi Mata or O Soto Gari) .
3. The rotational movements can enhance the throwing action changing the inner mechanics of Couple into Lever applying a Torque, changing the direction of one force during the attack or the time of application of one of the two couple forces.

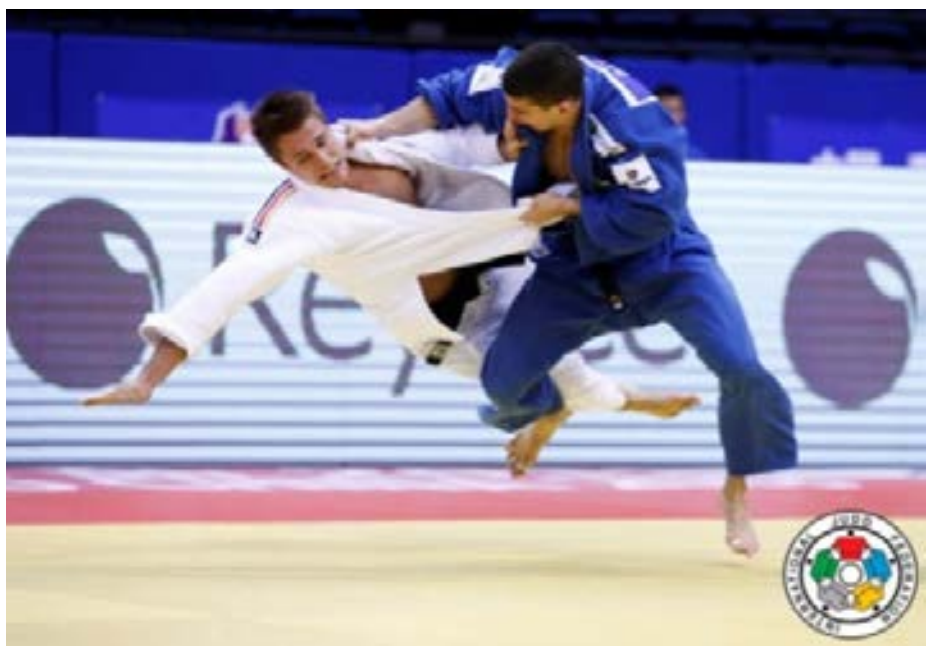

**Fig 20** ***Couple throws (O Soto Gari) Sideway application.***

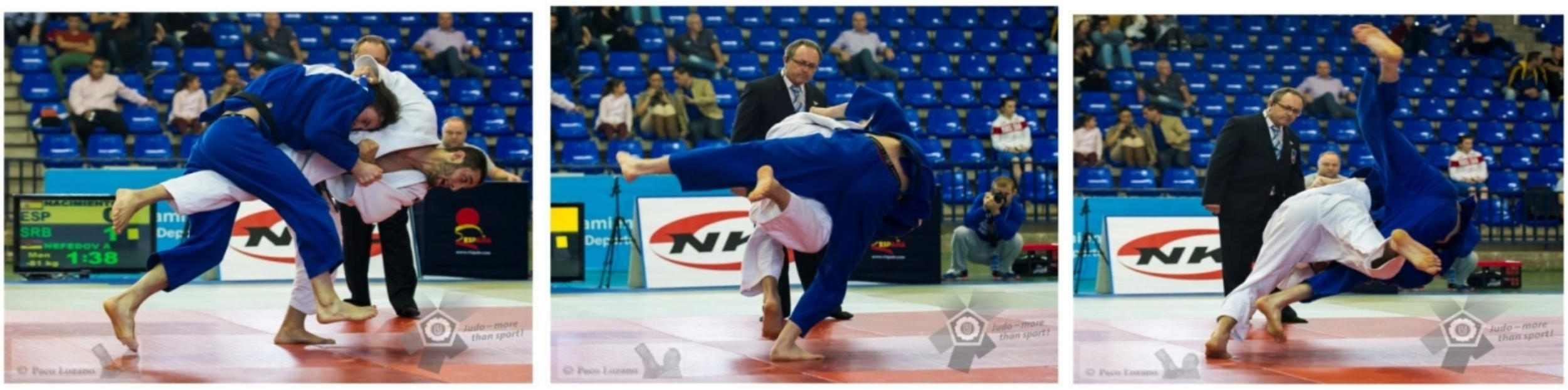

**Fig 21-23.** ***Couple throws (Uchi Mata) enhanced by a post rotation in the horizontal plane.***

Lever Techniques are enhanced in their effectiveness in four ways:

1. The rotational movements, strictly connected to the Lever techniques mechanics, achieving victory in competition, can be extended to the unbalance phase (Rotational Kuzushi)
2. The rotational movements can be applied in a totally new way putting away even the unbalance that is basic in the Lever techniques.
3. The Lever tool can be hybridized with the application of a Couple to lower the energy consumption and to overcome some strong defensive resistance.
4. The lifting action against gravity is an important tool connected to Lever that are friction dependent, with this action Tori detaches Uke's feet off the tatami by preventing any defensive action.

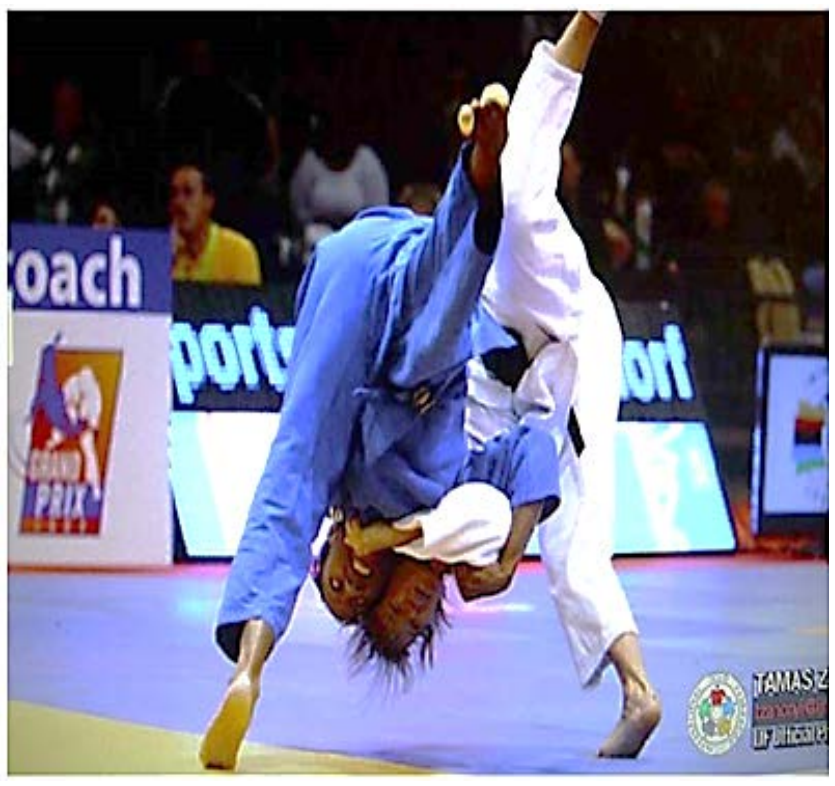

***Fig 24 Hybridization of Lever (Seoi Nage) with a Couple***

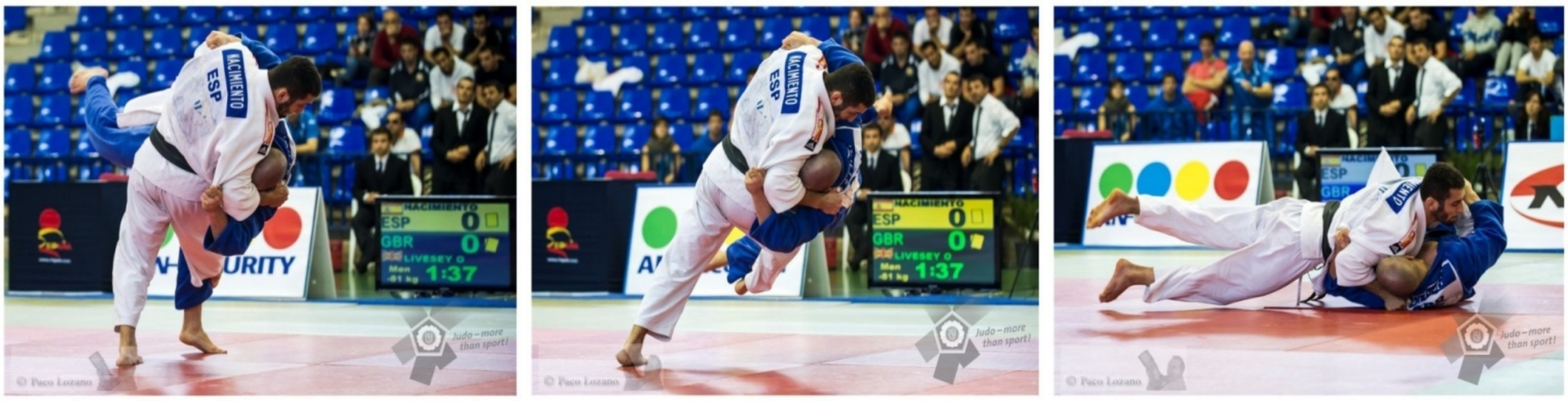

***Fig 25-27. Rotatory variation of Lever throw ( Hiza Guruma)***

Lifting against gravity is a powerful tool adopted in masterful way by Koga; the biomechanics that underlines this toll is to detach the feet off the mat nullifying friction between Use's feet and Tatami. Feet friction is the only way for Athletes in this case for Uke to transfer forces to Tory, utilizing ground reaction force and to resist to his Lever action applied.

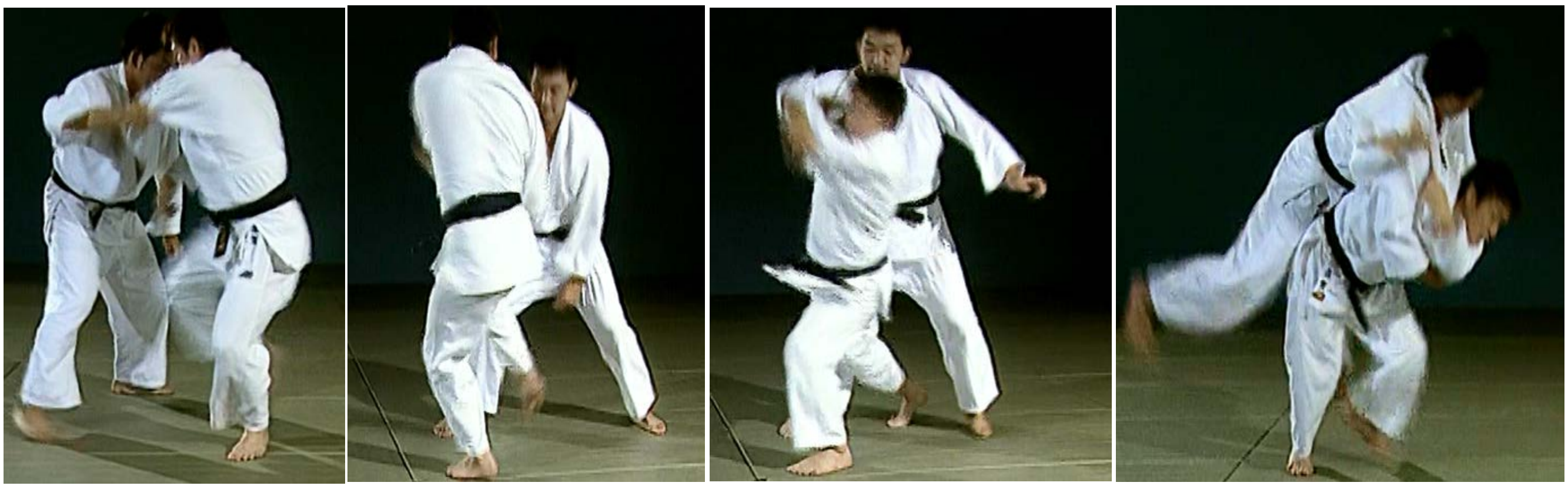

***Fig 28-31 The well known lift action Toshihiko Koga style:***
***1) Deep inside with the right leg 2) back with the left leg lifting up Uke's body and detaching his feet from the mat***

Without mat contact is impossible to apply any defensive action; very often today lifting is applied during suwari seoi action to perfect the throw.
***More widely lifting tool could be considered an overall tool useful and applicable during the first phase of every technique both Couple and Lever.***
However his effectiveness is greater in lever techniques, considering the mechanics of throws heavily depending from friction and unbalance.
See next figures in which this action is developed during a suwari seoi overcoming the Uke's defensive action and perfecting with Ippon the Lever throw.

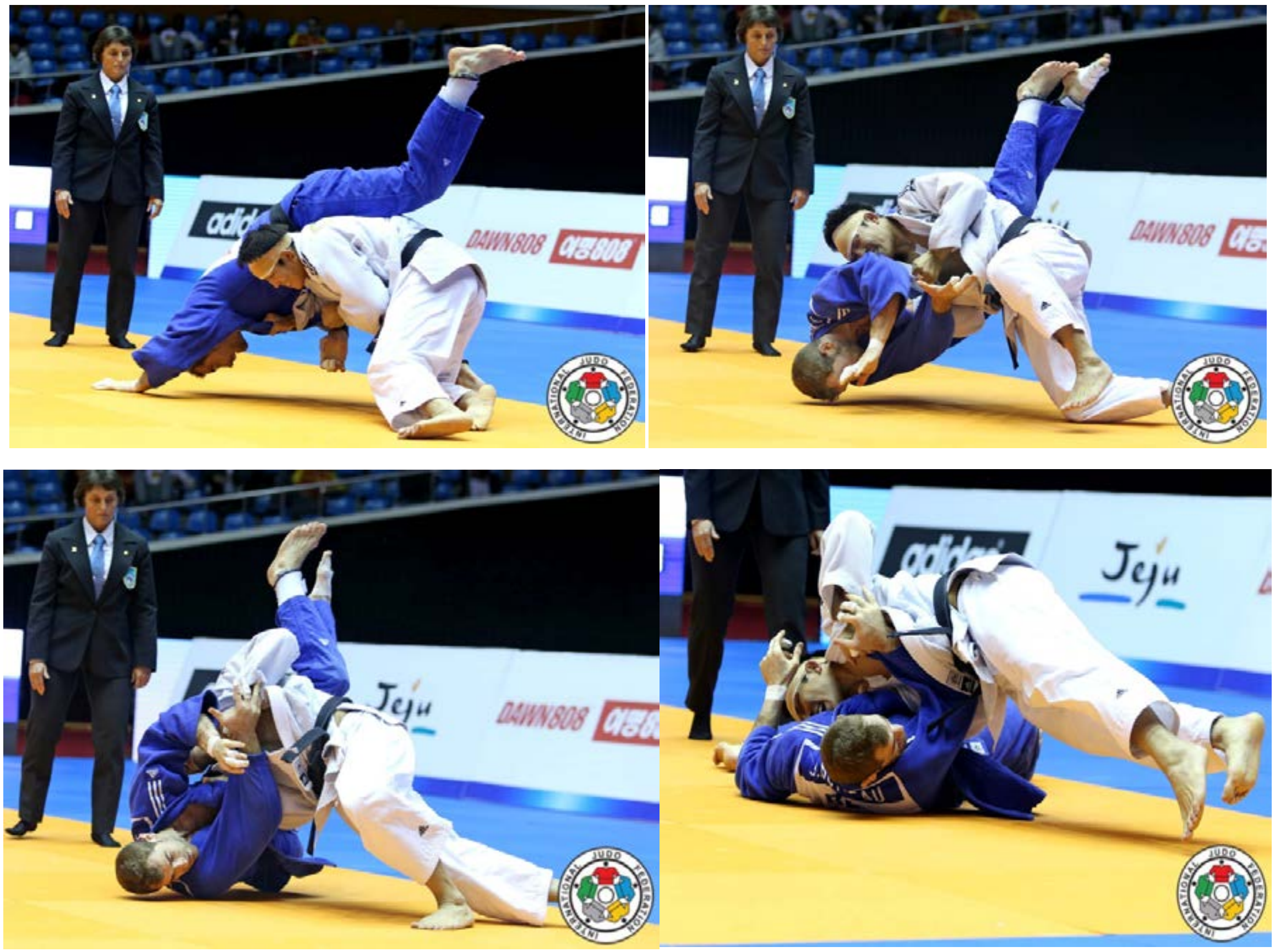

***Fig 32-35. Lifting action against gravity to perfect a Lever action (Suwari Seoi)***

## 6. Tools in combinations

Combination of throws is another tactical skill that usually is applied in high level judo competition.
How many combinations there are? Practically infinitive!
Unfortunately a clear justification to combine in effective way throwing techniques lacks in judo world.
Usually combinations are born from both study and fantasy of judo athletes and coaches.
Some are effective some others are useful only for exhibition, not applicable in the high dynamic situations that arise in competition.
Now a new question arises.
What are the combinations really effective in high level competition?
Biomechanics can help in finding the right way to combine throws in effective way, for high level competitions.
In fact it is possible to group all techniques in a new operative classification based on distance between athletes, and combinations can be built, considering the throwing techniques that could be assembled together in function of the variation of this distance.
The next step is to gather combination in function of throws' attack distance
This let us to find three specific groups:

**1) Chica Ma Waza, ( short distance applied techniques)**
**2) Ma Waza and ( medium distance applied techniques)**
**3) To Ma Waza. ( long distance applied techniques)**

These groups are both: in normal judo language and in biomechanical language, the basic blocks that could be connected to build all possible combinations.
Each group is collected without any specification about directions or applicative angles, which could be high variable.
In the next three tables are collected all useful techniques for one effective combination in high level competition.
Combinations are built connecting techniques of different groups for example:

Chica Ma Waza in connection with a Ma Waza → Uchi Mata connected to Tai Otoshi

→ Couple technique connected to a Lever technique

## *Combinative Classification of Throws*

## *(in function of attack distance)*

### *Tab.10 Chica Ma Waza (short distance applied techniques)*

(Tight body contact – applied by body rotation at zero/ low velocity and strong grips-)

| | |
|---|---|
| • Uki Goshi | • Lever |
| O Goshi | • |
| Seoi Nage | • |
| Koshi Guruma | • |
| O Guruma | • |
| Tsurikomi Goshi | • |
| Tsuri Goshi | • |
| All Innovative henka | • |
| All Chaotic throws | • |
| • Hane Goshi | • Couple |
| Harai Goshi | • |
| O Soto Gari sideway | • |
| Uchi Mata | • |
| All Innovative henka | • |
| Very few Chaotic throws | • |

## *Tab. 11 Ma Waza (medium distance applied techniques)*

(Medium distance applied with classical or double central grips)

| | |
|---|---|
| • Osoto Gari | Couple |
| Ouchi Gari | |
| All Innovative Henka | |
| • Sasae Tsurikomi Ashi | Lever |
| Harai Tsurikomi Ashi | |
| Uki Otoshi | |
| Sumi Otoshi | |
| O Soto Guruma | |
| Hiza Guruma | |
| Ashi Guruma | |
| Tai Otoshi | |
| All Innovative Henka | |
| All Chaotic Throws | |

## *Tab. 12. To Ma Waza (long distance applied techniques)*

(Applied at first contact, some are possible also with one sleeve grips only, or theoretically without grip, all are Couple group techniques)

| | |
|---|---|
| De Ashi harai/barai | Couple |
| Ko Soto gari/gake | |
| Ko Uchi gari/barai | |
| Okuri Ashi harai/barai | |
| Uchi Mata with one grip | |
| O Soto Gari sideway Jumping | |
| All Innovative Henka | |

**Biomechanical Principle to assemble Judo Combination**

***Combinations are produced modifying inter-athletes distance or type of bodies' contact changing both throwing technique and final direction, setting up a linear combination of the two tools*** **(Couple and Lever*)***

Biomechanical analysis helps to systematize combination connecting the two tools to throw to the shifting velocity of Athletes Couple System.

***V=0***

***Couple***

***Couple techniques ( To Ma, Ma and Chica Ma) Waza < Lever***

***Couple***

***Lever techniques ( Ma and Chica Ma) Waza < Lever***

***V= low/medium***

***Couple***

***Couple techniques (To Ma, Ma and Chica Ma) Waza < Lever***

***It is not possible apply Lever techniques at low/medium speed Tori needs to stop Uke, and then it come back to the first situation***

***V= high***

***Couple***

***Couple techniques ( To Ma Waza) < Lever***

***It is not possible apply Lever techniques at high speed Tori needs to stop Uke, and then it come back to the first situation.***

To underline the operative feature of the previous classification, there are pointed out some biomechanical remarks useful for coaching.
Generally speaking combinations for athletes specialized in To Ma Waza (long distance throws) that need timing, are organized on the basis of changing distance, from long one to shorter one (putting themselves closer to the adversary) with the support of right change of direction.
For athletes specialized in Chica Ma Waza (short distance applied throws) combination are grounded on changing direction: left/ right and vice versa or forward, backward, or forward, forward, or backward, backward. Right/ side, and backward /side changing are useful and effective but less frequent also in high level competitions.
Athletes preferring Ma Waza (medium distance applied throws ) can more freely applying both change direction and shortening distance. Obviously sutemi are closing combinations throws, in whatever directions.
Balance is essential in every combination and direction changing, both for Ma Waza and Chica Ma Waza, more often Tori and Uke are not well balanced on the first attack but balanced as Couple System by their dynamic balance.
Tori must fix with the unbalance action of his attack Uke body's position both on one or two legs in order to be able to change attack direction into the weak defensive side of Uke; this is possible when the unbalanced Uke, reacting to resist at a strong attack, becomes rigid and still.
Biomechanical analysis can upgrade the systematization of combination connecting the two tools to throws (Couple and Lever) to the shifting velocity of Athletes Couple System.
Biomechanics helps also, to clarify not only how to assemble combinations, but when to apply them in high level competition.
In effect a deep analysis of competitions let us to state the following points.

1. The only dynamical situation to build up complex combinations (3 or more techniques connecting Couple and Lever tools) is at zero shifting velocity of athletes system.
2. In such specific static situation, the fastest way to connect more techniques is to utilize the same one leg support position, changing goal at the moving "acting leg".
3. In term of "breaking symmetry" Tori must break Uke's symmetry stopping his body in unstable equilibrium on one leg, increasing his stability on it, in order to completely block his mobility.
4. The previous position will be assured if Tori add his own body weight on the stopped side of Uke's body.
5. Then in such situation change in direction of applied forces is only function of Tori's trunk rotation.
6. In theoretical way, combinations can be closed by every Sutemi, but as practical solution in high level competition that can't happen, because too dangerous, in fact in such high dynamical situation it is possible to mistake something and undergo an hold ( Osaekomi)
7. One Sutemi sometimes applied in combination during competition is Tani Otoshi, because the mechanics of technique brings a final position more safe for Tori.
8. More frequently is utilized as final combination tool , both from female and heavy weight categories, the Makikomi trick connected with Lever techniques

This Biomechanical theory of Combinations is grounded on a practical point of view, as the pragmatic western judo vision, today for one high performance coach is very important to give sound and easy understandable information to athletes improving their fighting capability.

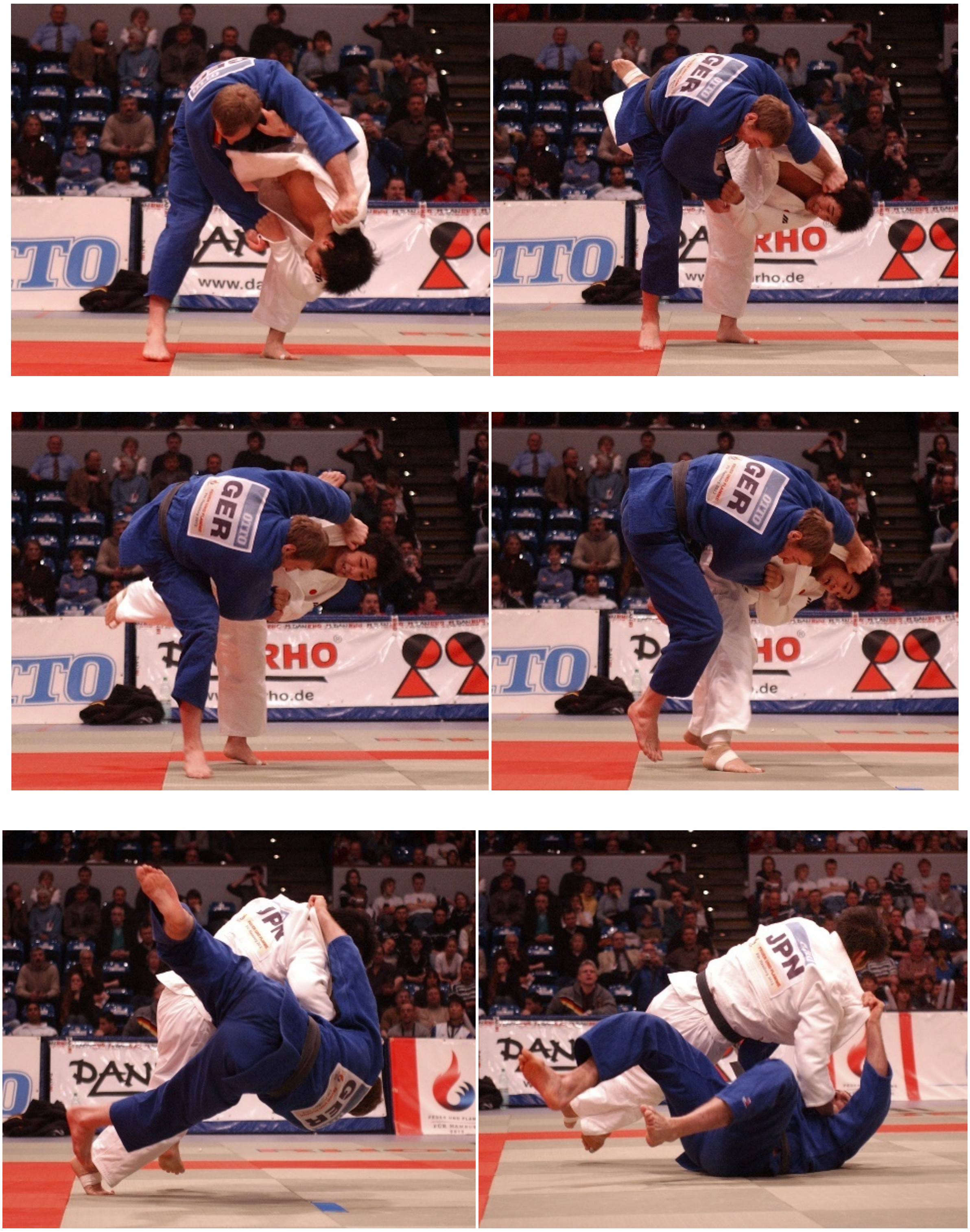

***Fig.36-41. Example of: Chica Ma Waza into Ma Waza, lengthening of distance with application of horizontal rotation and changing of Couple into Lever Tool.***

***(Uchi Mata into Sumi Otoshi)***

Inue against Hubert ( Finch)

## 7. Tools in action reaction attack

Many years ago Action-Reaction were connected and applied in the Judo world with a simple push/pull action, this action more often accomplished an instinctive reaction against the first action creating the right situation to apply the throw.
As regards to Action-Reaction tricks, due to the outstanding physical preparation and the increasing fighting experience of athletes it is practically impossible to apply them in high level competition, based only to simple push/pull action.
Today this trick must be integrated by a real technical attack often on a motionless Uke stopped on two feet by a breaking symmetry.
The biomechanical analysis of the Action-Reaction situations let us to determine some interesting finding:

1. Normally action reaction attack is finalized at only two techniques connected by opposite attack directions.
2. Also for the action reaction attacks, the most effective situation is at zero shifting velocity of athletes system.
3. In such specific static situation, the fastest way to connect two techniques is to utilize the same one leg support position, changing both direction of forces applied and goal at the “acting leg”.
4. The breaking symmetry frequently applied for this actions is to bend or turn Uke’s body stopped on his two feet.
5. The attack directions more often utilized are backward/ forward, forward/ backward, less frequent (but always possible) are left sideway/right sideway and vice versa.
6. In term of double central grip, the flexibility to apply double attack directly on the right and on the left by Ma Waza, is frequently connected with forward/ backward change of direction.

The Biomechanical theory of Action Reaction grounded on a practical point of view, as the pragmatic western vision of Initiative; it is simply connected to the application of two real attacks previously prepared in such way.
Tori applies a first real attack that can be stopped (in the better way) in only one direction and connects the first one to an effective second attack in this specific only direction, often there are connected two Couple techniques applied by the same leg, or a Couple and Lever the stopping point of the second one always applied by the same leg for saving attack time, making the action reaction more efficient and effective
Today for high performance coach is very important to prepare sound and easy understandable tactical tricks for the athletes to improve their fighting capability.
In the following figures there is one example of action reaction tactical tricks, applied by the same athlete that utilized a Couple-Lever as combination in the previous figures utilizing two Couple techniques backward –forward simply connected.

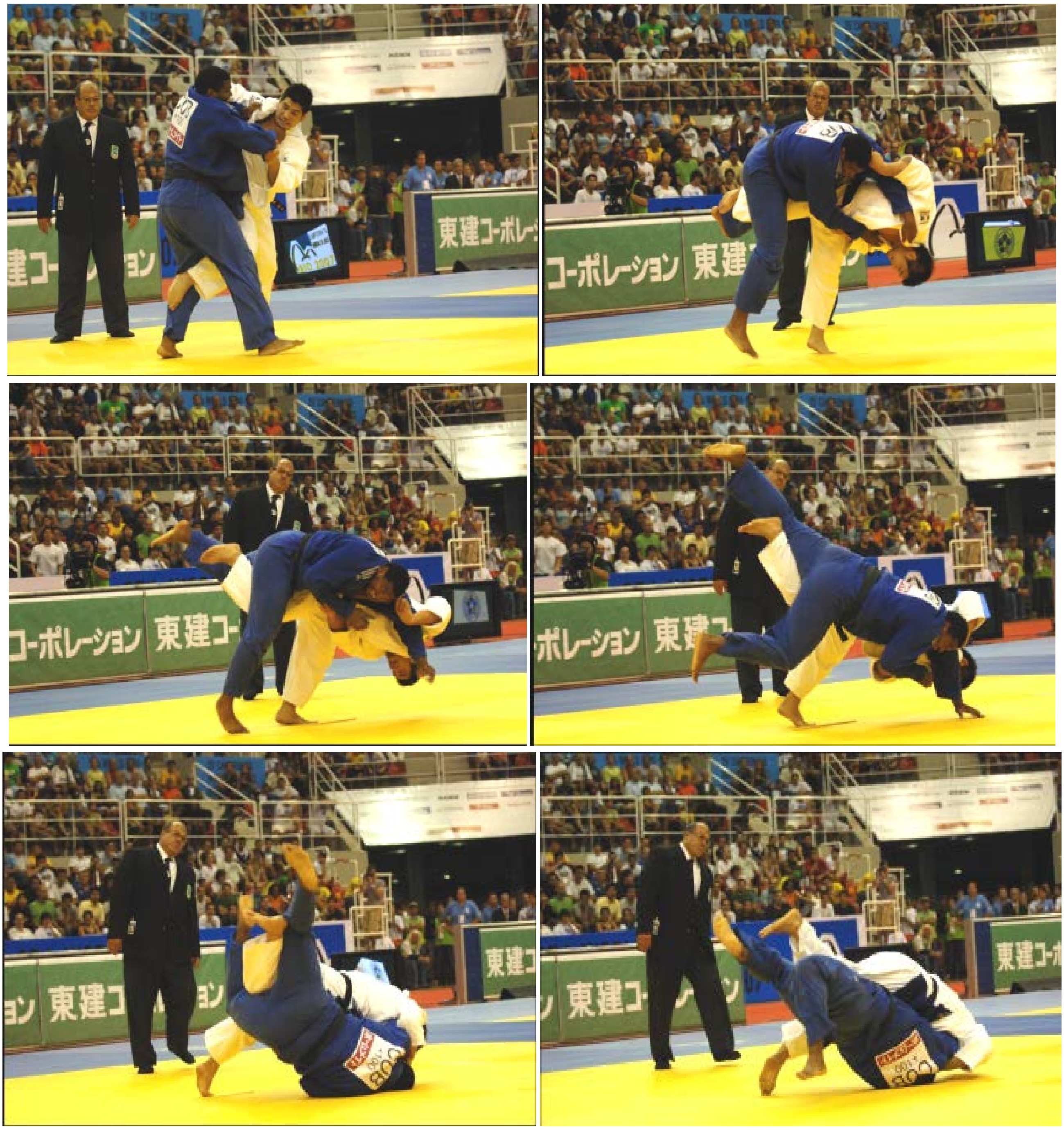

***Fig 42-47 Example of To Ma Waza into Ma Waza shortening of distance with application of backward/forward action-reaction on a natural simple reaction, using two Couple with the same leg.***

***(O Uchi Gari into Uchi Mata)***

Inue against Bryson ( Finch)

## 8. Physical and Biomechanical framework

How previously affirmed the Competition Model for Biomechanics is described by very simple words, Locomotion and Interaction, but the mathematics hidden in these words is very complex.
However the most important point in Biomechanics as in Physics is not to calculate the numerical value of specific phenomenon, but to single out the general mechanical law (as equation) that rules all similar situations.

### *Locomotion*

Into the Couple, human bodies of Athletes show complex responses connected to the human physiology whereof Brownian motion, starting from fractals till to multi-fractals aspects, is one of the basic modeling [ 45, 46, 47 ]. If we study, in deep, the change of position in time of bodies in space, starting from the motion of Centre of Mass in standing quiet position, till to Couple of Athletes ground track motion, Brownian Dynamics shows his ubiquitous presence in the motion description of these "Situation Sport"[48] If we consider the couple of athletes as a single system, then the motion of the centre of mass system is definite by a push pull random forces, which are directly connected to the friction forces between feet and mat.
This is the biomechanical base of the known judo paradox that state: ***"the most important part of the grip is the feet position".***
The whole system is isolated, no external forces, less the random push-pull forces by the friction force, then the fight general motion equation will be a Langevin type equation, (Sacripanti 1992) and it will be possible to write [49]:

$$F = \dot{v} = -\frac{\mu}{m}v + \frac{u}{m}\sum_j (\pm 1)_j \delta(t - t_j) = F_a + F' \qquad (1)$$

The experimental proof of this model can be founded in one Japanese work, on the world championship of the 1971. [18]
In the next figure from the Japanese work, we can see the summation of motion patterns of 1,2,7, and 12 judo fights; it is easy to understand that the random fluctuation not have a preferential direction over the time. In mathematical terms this means that $\langle F' \rangle = 0$ (2)
And from that, it is possible to assert that the motion of the Centre of Mass of the systems is Brownian.

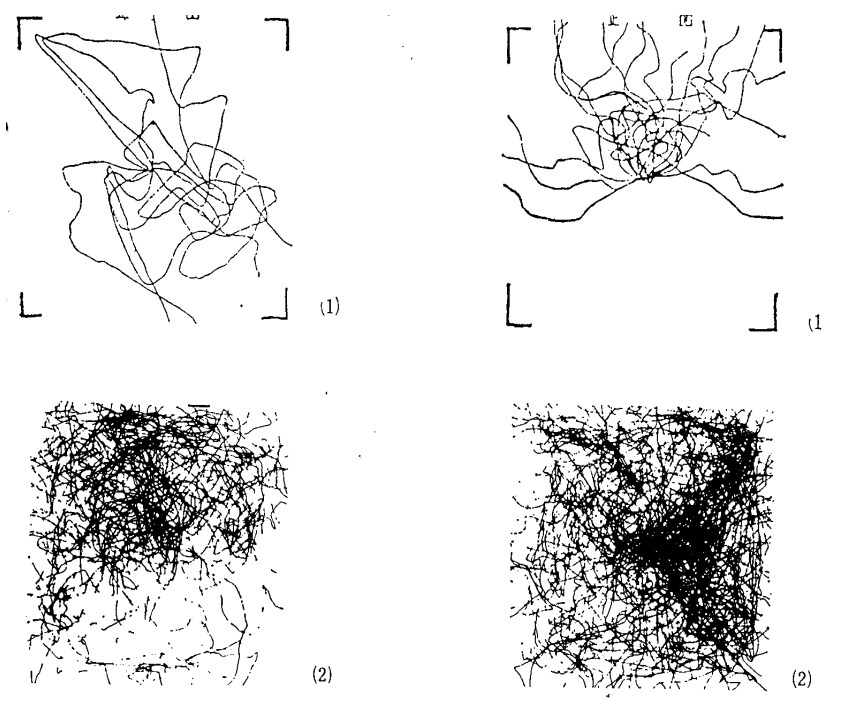

$\rightarrow \langle F' \rangle = 0$ [18]

***Fig.48. Summation of motion patterns of 1, 2, 7, and 12 fights.*** [18]

Numerical evaluation gives a confirmation of the previous Hypothesis validated by experimental data

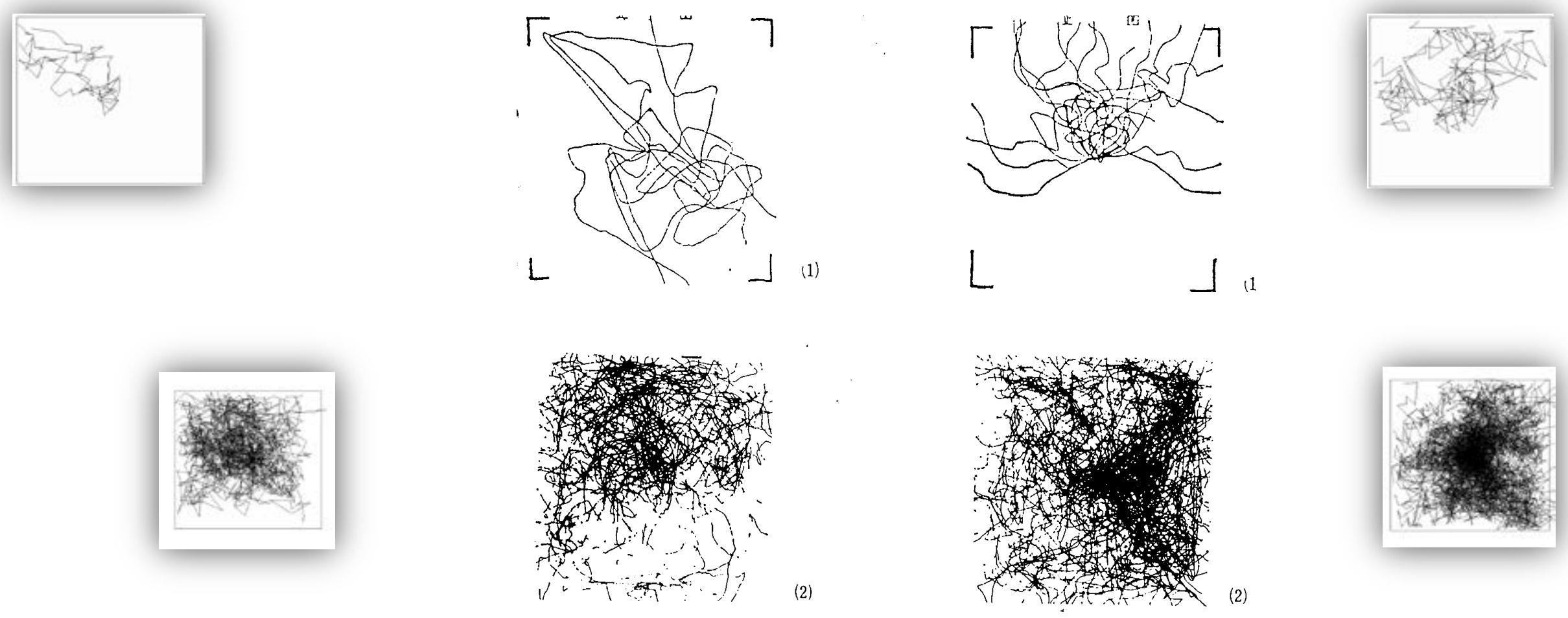

***Fig 49. Numerical reconstruction of Brownian Paths.*** [50]

## *Interaction*

***Dynamics of a body in a field of elastic forces with clockwise and counter-clockwise rotation***

The two Bodies that play judo are mainly connected by arms that apply push pull forces in every direction. In this sense we can compare the body drifting apart or getting close at a situation of dynamic of a body in a field of elastic internal forces. The Bertrand theorem assure us that the only closed orbit for the elastic generalized force

$F = -kx^{\alpha}$ (3)

There are only for α = 1; or α= -2 as already demonstrate in [51].

Interesting it is to analyze some of the trajectories (GAI) that Tori normally use to apply some lever techniques or some couple techniques.

Most Chica Ma Waza , few Ma Waza and no To Ma Waza. are dependent by inward rotation.

How it easy to see in the next figure all trajectories are similar to it, because Tori taking firmly a contact point performs a fast inward rotation to apply the lever or couple and throws Uke.

It is interesting that always speaking in term of elastic field, the class of trajectories applied by Tori are similar to a capture trajectory of a particle by another in elastic field with internal forces

$F_{ab} = -F_{ba} =$ F (4)

In this case the two particles act like the two athletes bodies during the inward rotation throw. [52]

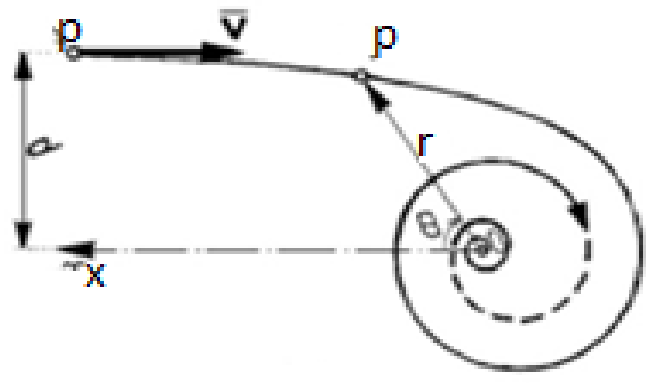

***Fig .50. Trajectory of a capture particle similar to Tori's throwing inward trajectory in judo interaction*** [52]

## *Almost-Plastic collision of extended bodies.*

Most often the end of the previous trajectory flows into a projection by a lever techniques that starts with a collision, which can be considered an almost plastic collision because the two athletes are strictly connected together, but obviously their bodies are not merged, then it is a collision almost-plastic of two extended bodies. In this case considering both athletes of more or less equal mass, the equations are very easy to obtain. [53]
If they have different starting velocities, after the contact they move connected till to the fall down that can be considered as a free fall.
In this case considering always negligible the gravity force, that increases greater and greater his importance into the motion after the collision, till to landing of Uke body (free fall); it is possible to write for the early instants of the collision, remembering that is a rotational impact: [54]

$$mv_1 + mv_2 = 2mv \qquad (5)$$

$$\textit{the conservation of angular moment give us } I_1\omega_1 = (I_1 + I_2)\omega_f \qquad (6)$$

$$or \ \frac{\omega_f}{\omega_1} = \frac{I_1}{(I_1 + I_2)} \qquad (7)$$

$$\textit{the impact is totally inelastic, and the loss of kinetic Energy is:}$$

$$\Delta K = \frac{1}{2} I_1\omega_1^2 - \frac{1}{2}(I_1 + I_2)\omega_f^2 = \frac{1}{2}\frac{I_1 I_2}{I_1 + I_2}\omega_1^2 \qquad (8)$$

## *Full rotation with free fall*

Complex rotational application of judo throwing techniques that will be analyzed in this section are connected to the tactics of direct attack applying a lever techniques with a fast and complete inward rotation, like spinning top at variable rotational inertia or vortex
It is, for Tori as observer, a clear study of Forces applied in a rotating reference frame.
At first it is important to evaluate the velocity transformation formula from the inertial to the rotating frame, this means how the speed is evaluated by people (public as observer) and Tori as observer during the execution of a rotating throws, as already demonstrate in a previous paper [55] the result is:

$$v = V + \left(\frac{dr'}{dt}\right)_O = V + (v' + \omega \wedge r') \qquad (9)$$

It is important to evaluate the general equation of motion of this variable mass spinning up, remembering the classical Newton approach to the rotational dynamics, [56] we can write:
The torque on the first athlete $\tau$ will be

$$\tau = r\,F \ (10)$$

Where $r$ is the radius between the centre of mass and the point where the force $F$ is applied.
In term of rotational dynamics this equation can be written also as:

$$\tau = I\frac{d\omega}{dt} = mr^2\frac{d\omega}{dt} \qquad (11)$$

It is very easy to solve this equation if the athlete is up-righted like a symmetric spinning top, because as long as the torque is applied in such a way as to increase (or decrease) its rotational speed around the $\hat{z}$-axis, this is just a one-dimensional equation and offers no surprises. [57]

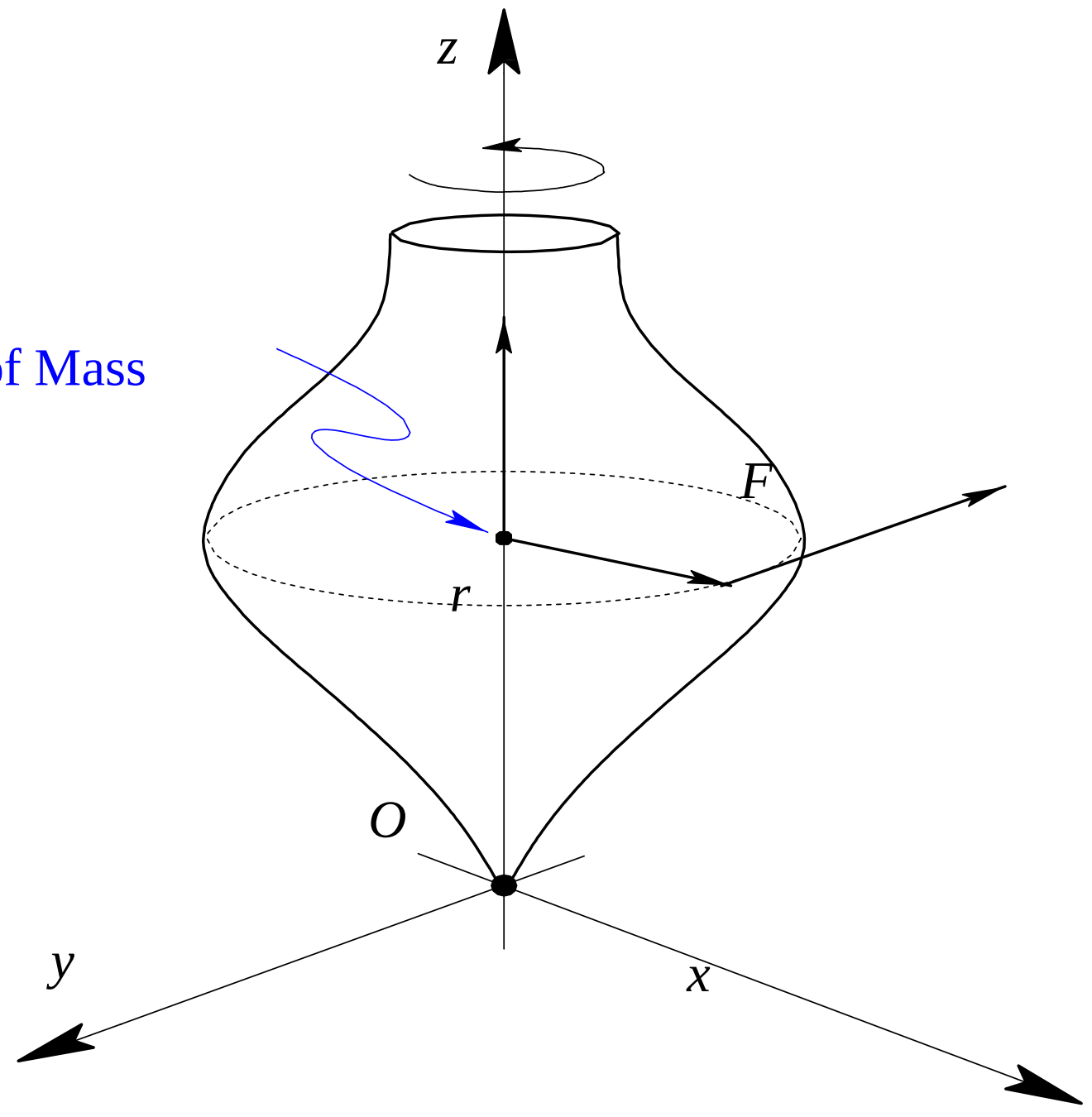

***Fig. 51 Spinning Top***

It is necessary to remember that if we use the Euler representation of a rotating rigid body they produce a non linear system of equation not always resolvable.
However interesting are the equations of motion of the Euler angle found by Garanin [58], which are shown in the next:

$$\dot{\theta} = \left(\frac{1}{I_1} - \frac{1}{I_2}\right) L \sin\theta \sin\psi \cos\psi$$

$$\dot{\varphi} = \left(\frac{\sin^2\psi}{I_1} + \frac{\cos^2\psi}{I_1}\right) L \qquad (12)$$

$$\dot{\psi} = \left(\frac{1}{I_3} - \frac{\sin^2\psi}{I_1} - \frac{\cos^2\psi}{I_1}\right) L \cos\theta$$

Note that equations for $\dot{\theta}$ (nutation) and $\dot{\psi}$ (spin) form an autonomous system of equations that can be solved as first step, after that the equation for the precession $\dot{\varphi}$ can be integrated using the previous found spin, obtaining the demanded solution.

If, however, athlete ***tilts*** [59] his rotational axis through an angle $\varphi$, as shown in Figure 40, the situation gets a little more complicated and a lot more interesting respect to a spinning top.
In the case of the tilted athlete shown in Figure 41, gravity pulls down on the centre of mass of the athlete, which would pull a non-spinning athlete (because it is in unstable equilibrium) downward and simply increase the tilt angle $\varphi$ as the athlete falls down.
Normally in a spinning top the torque, and thus the change in the angular-momentum vector, is perpendicular to the axis $\hat{u}$, which leads the top to move “sideways” in a circle around the *z*-axis, and this motion is called *precession,* but this well known phenomenon is nullified in our case by the increased mass of the system that after the (quasi-plastic collision) firmly connect together the two athletes bodies.

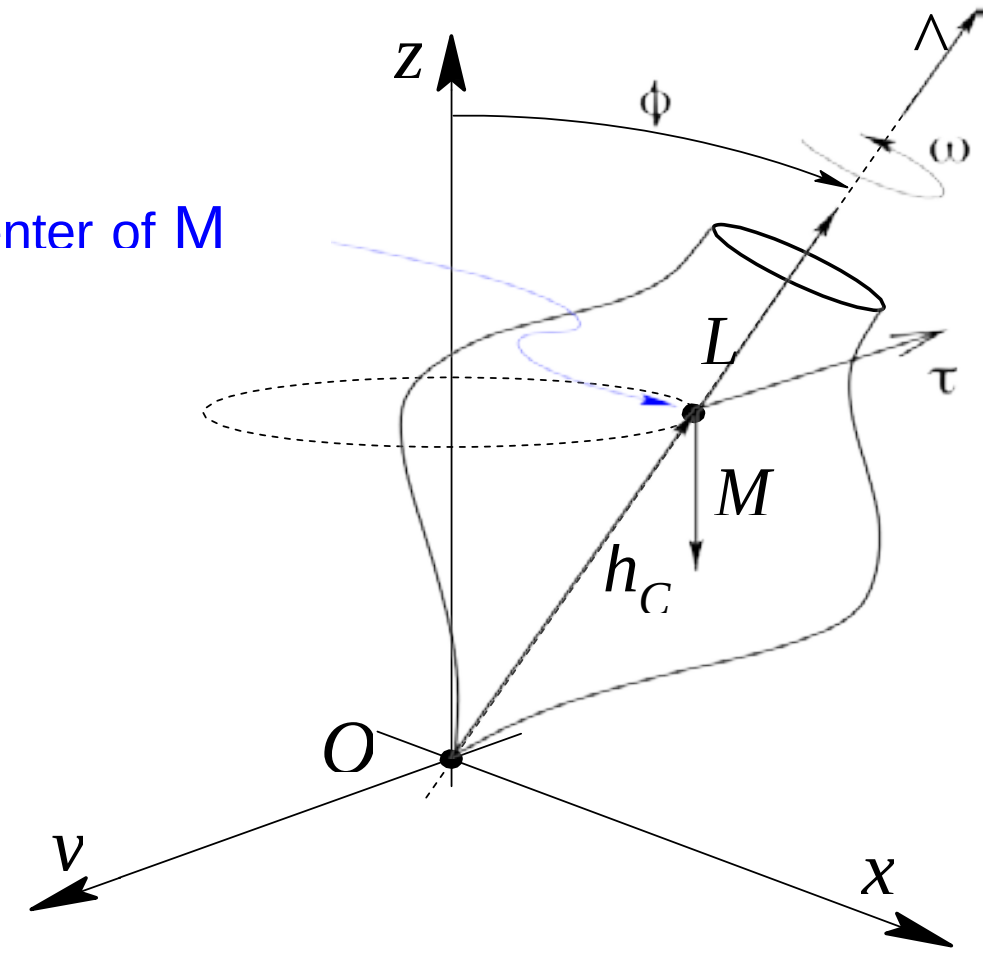

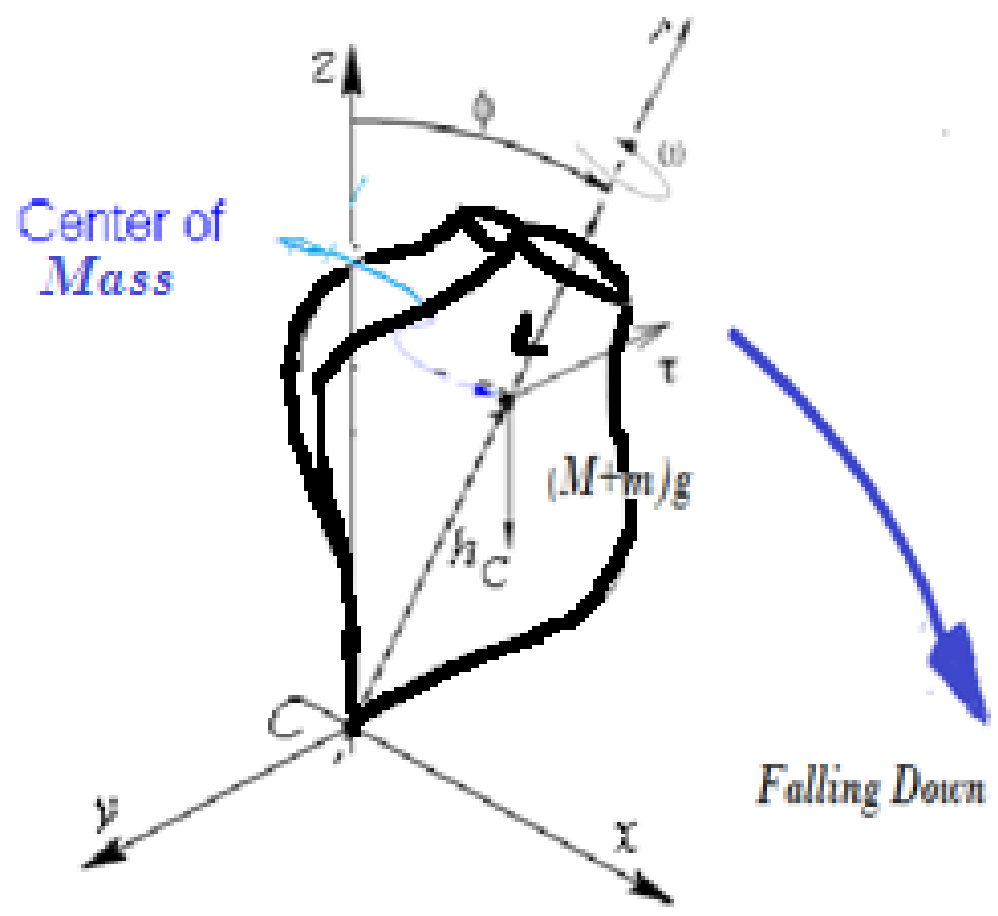

***Fig.52 .Tilting Spinning Top;*** ***Fig. 53. Variable Inertia Falling Spinning Top.***

The three - dimensional equation becomes:

$\tau = r \wedge F$ *[16]*

$$\tau = \frac{d(I\omega)}{dt} = \frac{d[(M+m)\omega]}{dt} \qquad (17)$$

Abruptly after contact, mass more or less double, like a system at variable mass,[60] velocity drops down and external gravity force, overcoming the potential precession motion, helps the bodies to fall down.
For Lever throws tactical tools are or rotational applications or hybridization by a supplementary torque
For the tactics of Couple throws there is on the application of a linear combination of torques in different symmetry planes.
For Couple throws all tactical tools applied are always torques.

## 9. Conclusion

In this Invited paper we are faced with the problem of enhancement of judo throwing techniques.
Many Biomechanics and Physiological studies shown, that Couple techniques are not only more efficient than Lever techniques but also energetically more convenient.
Tactics in high level competition is classified by researches in the world by the triad: Direct Attack, Combination Attack and Action-Reaction Attack.
For direct attack the enhancement is grounded for lever techniques mainly on the introduction of rotational approach or on the hybridization of lever with couple movements, and makikomi attack
For couple techniques mainly by the sideway attack and by the introduction of a post rotation in the horizontal plane and some time by the shifting of couple in lever changing the direction of forces.
Lifting up action could be useful in few Couple techniques, and in all Lever techniques both: as nullifying friction tool, and as perfecting action in the final kake phase, for this aspect is very useful in all drop variation of Lever techniques applied, specifically in the Suwari Seoi throw.
For combinations and action-reaction attack we must underline that in term of effectiveness in high level competition ***fast combination*** or ***fast action reaction*** application are preferable.
These specific applications are characterized by some fixed aspects.
Couple of Athletes system fixed, Tori attacks continuously by techniques with one support leg and one acting leg, Uke that undergoes the attack is on a leg support during combination attacks, and two legs support on action reaction attacks.
The fast shifting in a “new techniques” for Tori is simply connected to the change in action of “acting leg” ( from sweep to stopping point, or from sweep in one direction to another direction, etc.).
These actions that are very fast, by their mechanical aspects, are very difficult to avoid by reason of their shortest time of execution.
These two specific tactical attack ways are the preferred and most utilized in high level competitions as combinations or action reaction actions.
From the other hand, for teachers that prefer Japanese approach to Judo, Biomechanics demonstrated clearly that ***Action Reaction attack*** is a little subset of ***Combination*** and not a separate way to apply tactics in competition.
Then in Japanese way if ***Direct Attack*** flows in Renzoku Waza, ***Combination*** flows in Renraku Waza.
***Action Reaction Attack*** is a specific subset of Renraku Waza, with the limitation that the second throws must have the opposite direction of the first one.

## 10.References